\documentclass[sigplan,10pt]{acmart}
\renewcommand\footnotetextcopyrightpermission[1]{}

\def\buildextended{}  %

\usepackage{times}
\usepackage{float} %
\usepackage{bm} %

\usepackage{multicol}
\usepackage{booktabs}  %
\usepackage{dcolumn}   %
\usepackage{multirow}  %
\usepackage{rotating}
\usepackage{xspace}
\usepackage{hyphenat}  %
\usepackage{enumerate}
\usepackage{enumitem}
\usepackage{mfirstuc} %
\usepackage{tikz} %
\usepackage{xparse} %
\usepackage{listings}
\definecolor{dkgreen}{rgb}{0,.6,0}
\definecolor{dkblue}{rgb}{0,0,.6}
\definecolor{dkyellow}{cmyk}{0,0,.8,.3}
\usepackage{wrapfig}
\usepackage{textcomp}
\usepackage{tabularx}
\usepackage{pifont}
\usepackage{url}

\usepackage{breakurl}

\usepackage{wrapfig}

\usepackage{ulem}
\usepackage{mfirstuc}

\usepackage{colortbl} %
\usepackage{array}    %

\usepackage{dblfloatfix}

\usepackage[font=small]{caption}
\usepackage{subcaption}

\usepackage[noend]{algpseudocode}
\algrenewcomment[1]{\hfill// #1}%
\algrenewcommand\algorithmicindent{1.25em}

\definecolor{darkgreen}{rgb}{0,0.75,0}

\algrenewcommand\algorithmicdo{:}
\algrenewcommand\algorithmicwhile{\textbf{while}}
\algrenewcommand\algorithmicfor{\textbf{for}}
\algrenewcommand\algorithmicforall{\textbf{for all}}
\algrenewcommand\algorithmicloop{\textbf{loop}}
\algrenewcommand\algorithmicrepeat{\textbf{repeat}}
\algrenewcommand\algorithmicuntil{\textbf{until}}
\algrenewcommand\algorithmicprocedure{\textbf{procedure}}
\algrenewcommand\algorithmicfunction{\textbf{function}}
\algrenewcommand\algorithmicif{\textbf{if}}
\algrenewcommand\algorithmicthen{:}
\algrenewcommand\algorithmicelse{\textbf{else}:}
\algrenewcommand\algorithmicrequire{\textbf{Require}:}
\algrenewcommand\algorithmicensure{\textbf{Ensure}:}
\algrenewcommand\algorithmicreturn{\textbf{return}}
\algdef{SE}[WHILE]{While}{EndWhile}[1]{\algorithmicwhile\ #1\algorithmicdo}{\algorithmicend\ \algorithmicwhile}%
\algdef{SE}[FOR]{For}{EndFor}[1]{\algorithmicfor\ #1\algorithmicdo}{\algorithmicend\ \algorithmicfor}%
\algdef{S}[FOR]{ForAll}[1]{\algorithmicforall\ #1\algorithmicdo}%
\algdef{SE}[IF]{If}{EndIf}[1]{\algorithmicif\ #1\algorithmicthen}{\algorithmicend\ \algorithmicif}%
\algdef{C}[IF]{IF}{ElsIf}[1]{\algorithmicelse\ \algorithmicif\ #1\algorithmicthen}%
\algnotext{Function}
\algnotext{EndFunction}
\algnotext{EndFor}
\algnotext{EndIf}
\algnotext{EndWhile}
\algnewcommand\Or{\textbf{or}\xspace}
\algnewcommand\myAnd{\textbf{and}\xspace}

\usepackage{amsmath,amssymb,amsthm}

\usepackage[noeka]{mathrmletter}
\usepackage{tgtermes}
\usepackage{afterpage}
\newif\ifextended
\newif\iflongbatching
\newif\ifsubmission
\newif\ifelementary
\newif\ifwithdbproof
\newif\ifbuildanonapdx

\ifx\buildextended\undefined
\else
  \extendedtrue
\fi
\def\noeditingmarks{}

\def\hn{\sffamily\selectfont}
\newcommand{\mpfont}{\hn\scriptsize}

\ifx\noeditingmarks\undefined
  \newcommand{\MPworker}[2]{\unskip{\color{#1}\vrule\vrule}{\marginpar{\raggedright\color{#1}\mpfont #2}}}
  \newcommand{\pgwrapper}[3]{\begingroup \color{#1} #2: #3 \endgroup}
  
  \newcommand{\pgwrapperb}[1]{\textbf{#1}}
  \newcommand{\dangerwrapper}[1]{{\color{red}#1}}
  
\else
  \newcommand{\MPworker}[2]{\unskip}
  
  \newcommand{\pgwrapperb}[1]{}
  \newcommand{\pgwrapper}[3]{}
  \newcommand{\dangerwrapper}[1]{}

\fi

\newcommand{\jian}[1]{\pgwrapper{purple}{JZ}{#1}}
\newcommand{\cheng}[1]{\pgwrapper{blue}{CT}{#1}}

\newcommand{\CP}[1]{\MPworker{blue}{CT: #1}}

\def\t{\textit}

\newcommand{\CF}[1]{\xmakefirstuc{#1}}

\newcommand{\sys}{\textsc{Boomslang}\xspace}
\newcommand{\Sys}{\CF{\textsc{Boomslang}}\xspace}
\newcommand{\Port}{Port\xspace}

\newcommand{\circledone}{\ding{192}\xspace}
\newcommand{\circledtwo}{\ding{193}\xspace}
\newcommand{\circledthree}{\ding{194}\xspace}

\makeatletter
\def\imod#1{\allowbreak\mkern10mu({\operator@font mod}\,\,#1)}
\makeatother


\def\compactify{\itemsep=0in \topsep=2pt \parsep=0.00in \partopsep=0pt
  \leftmargin=2em}
\let\latexusecounter=\usecounter

{\begin{list}{\labelitemi}{\itemsep3pt \topsep3pt \parsep0.00in
          \partopsep=3pt \leftmargin1em}}%
          {\end{list}}
\newenvironment{myitemize2}%
{\begin{list}{\labelitemi}{\itemsep1pt \topsep2pt \parsep0.00in
          \partopsep=0pt \leftmargin1.2em}}%
          {\end{list}}
{\begin{list}{\labelitemi}{\itemsep1pt \topsep2pt \parsep0.00in
          \partopsep=0pt \leftmargin2em}}%
          {\end{list}}
{\begin{list}{\labelitemi}{\itemsep2pt \topsep2pt \parsep0.00in
          \partopsep=0pt \leftmargin1.2em}}%
          {\end{list}}
{\begin{list}{\labelitemi}{\itemsep3pt \topsep3pt \parsep0.00in
          \partopsep=3pt \leftmargin1.5em}}%
          {\end{list}}
{\begin{list}{\S3.3}{\itemsep3pt \topsep3pt \parsep0.00in
          \partopsep=3pt\leftmargin1em}}%
          {\end{list}}

\def\compactsortof{\itemsep=0in \topsep=2pt \parsep=0.00in \partopsep=0pt
  \leftmargin=1.2em}
\newenvironment{myenumerate2}
{\def\usecounter{\compactsortof\latexusecounter}
  \begin{enumerate}}
    {\end{enumerate}\let\usecounter=\latexusecounter}

\def\compactsqueeze{\itemsep=0pt \topsep0pt \parsep=0ex \partopsep=0pt
  \leftmargin=1.63em}

\ifextended
  \def\compactRenum{\itemsep=1ex \topsep=1ex \parsep=0.00in \partopsep=0pt
    \leftmargin=2.05em}
\else
  \def\compactRenum{\itemsep=0in \topsep=2pt \parsep=0.00in \partopsep=0pt
    \leftmargin=2.05em}
\fi

\ifx\normalpar\undefined
  
\else
  
\fi

\def\discretionaryslash{\discretionary{/}{}{/}}
{\catcode`\/\active
\gdef\URLprepare{\catcode`\/\active\let/\discretionaryslash
\def~{\char`\~}}}%
\def\URL{\bgroup\URLprepare\realURL}%
\def\realURL#1{\tt #1\egroup}%

\NewDocumentCommand{\xrightarrows}{ O{}O{} }{%
\mathrel{%
  \vcenter{\hbox{%
      \begin{tikzpicture}
        \node[minimum width=0.5cm,minimum height=1ex,anchor=south,align=center] (a){\text{\vphantom{hg}#1}\\[0.5ex] \vphantom{hg}#2};
        \draw[->] ([yshift=0.25ex]a.west) -- ([yshift=0.25ex]a.east);
        \draw[<-] ([yshift=-0.25ex]a.west) -- ([yshift=-0.25ex]a.east);
      \end{tikzpicture}
    }}%
}%
}

\newcommand{\wwarrow}[1]{\xrightarrow[]{\text{ww(#1)}}}
\newcommand{\wrarrow}[1]{\xrightarrow[]{\text{wr(#1)}}}
\newcommand{\rwarrow}[1]{\xrightarrow[]{\text{rw(#1)}}}

\newcommand{\wwarrowx}{\xrightarrow[]{\text{\textit{ww}}}}
\newcommand{\wrarrowx}{\xrightarrow[]{\text{\textit{wr}}}}
\newcommand{\rwarrowx}{\xrightarrow[]{\text{\textit{rw}}}}
\newcommand{\pwrarrowx}{\xrightarrow[]{\text{\textit{pwr}}}}
\newcommand{\prwarrowx}{\xrightarrow[]{\text{\textit{prw}}}}

\newcommand{\heading}[1]{
  \vspace{1ex}
  \noindent
  \textbf{#1}}
\newcommand{\headingzero}[1]{
  \noindent
  \textbf{#1}}

\newcommand{\ttt}{\texttt}

\def\t{\textit}

\theoremstyle{acmdefinition}

\newcommand{\iso}{isolation level\xspace}

\newcommand{\ser}{serializability\xspace}
\newcommand{\Ser}{Serializability\xspace}
\newcommand{\si}{SI\xspace}

\newcommand{\checker}{checker\xspace}
\newcommand{\checkers}{checkers\xspace}
\newcommand{\blindwt}{BlindW-T\xspace}
\newcommand{\trace}{trace\xspace}
\newcommand{\traces}{traces\xspace}

\newcommand{\obss}{observed events\xspace}

\newcommand{\specs}{specifications\xspace}

\newcommand{\unknowns}{unknowns\xspace}

\newcommand{\posb}{possibility\xspace}
\newcommand{\posbs}{possibilities\xspace}
\newcommand{\superpos}{superposition\xspace}
\newcommand{\irimply}{implication\xspace}
\newcommand{\irimplies}{implications\xspace}
\newcommand{\superposs}{superpositions\xspace}
\newcommand{\Superpos}{Superposition\xspace}

\newcommand{\unsatsearch}{unsat search\xspace}

\newcommand{\asg}{Abstract Semantic Graph\xspace}

\newcommand{\vorder}{version order\xspace}
\newcommand{\tombstone}{tombstone\xspace}
\newcommand{\tombstones}{tombstones\xspace}

\newcommand{\uval}{unique value\xspace}
\newcommand{\uvals}{unique values\xspace}

\newcommand{\truedeletions}{physical deletions\xspace}
\newcommand{\tval}{non-unique values\xspace}
\newcommand{\mnull}{\texttt{NULL}\xspace}

\newcommand{\wrrel}{read-dependency\xspace}
\newcommand{\wrrels}{read-dependencies\xspace}
\newcommand{\wwrel}{write-dependency\xspace}
\newcommand{\wwrels}{write-dependencies\xspace}
\newcommand{\rwrel}{anti-dependency\xspace}
\newcommand{\rwrels}{anti-dependencies\xspace}
\newcommand{\pwrrel}{predicate-read-dependency\xspace}

\newcommand{\prwrel}{predicate-anti-dependency\xspace}
\newcommand{\prwrels}{predicate-anti-dependencies\xspace}

\newcommand{\iter}{iterator\xspace}
\newcommand{\iters}{iterators\xspace}

\newcommand{\optone}{topological prioritization\xspace}
\newcommand{\opttwo}{reachability pruning\xspace}

\newcommand{\readinfo}{read lineage\xspace}
\newcommand{\writeinfo}{write history\xspace}

\newcommand{\varwset}{\t{wSet}}
\newcommand{\varvo}{\t{verOrder}}
\newcommand{\varpotentialw}{\t{candWs}}

\setcopyright{none}

\begin{document}
\setlength{\emergencystretch}{3em} %

\title{ \huge
  Making Transaction Isolation Checking Practical
}

\author{\textup{Jian Zhang$^\star$, Shuai Mu$^\ddagger$, and Cheng Tan$^\star$}\\
\fontsize{9.5}{11}\selectfont
\textup{$^{\star}$Northeastern University
    \quad\quad\quad 
    $^{\ddagger}$Stony Brook University}}

\begin{abstract}

Checking whether database transactions adhere to isolation levels is a crucial yet challenging problem.
We present \sys, the first general-purpose checking framework
capable of verifying configurations that were previously uncheckable.
\sys advances beyond prior work in three key aspects:
    (1) it supports arbitrary operation types provided by modern transactional key-value stores,
    (2) it requires no knowledge of database internals,
    and (3) it offers a modular, extensible pipeline amenable to customization and optimizations.

\sys adopts a front-/back-end separation.
As the front-end,
it parses a database trace into an \emph{\asg},
    which is then lowered---via semantic analysis---into a low-level intermediate representation (IR).
The back-end converts this IR to a set of constraints for SMT solving.
This design is enabled by a key abstraction in the IR,
    called \emph{\superposs}, which
    capture the uncertainty and complexity caused by
    arbitrary operations and missing information.
Our experiments show that with just 271--386
    lines of code,
    the core logic of three prior checkers can be reimplemented as \sys modules,
    achieving comparable or superior performance.
Using \sys, we also
    identify a new bug in TiDB,
    audit the metadata layer of the JuiceFS file system,
    check vendor-specific behavior in MariaDB,
    support five previously unchecked isolation levels,
    and confirm a theoretical result on the correctness of strict serializability.

\end{abstract}
 
\settopmatter{printfolios=true, printacmref=false}
\maketitle
\pagestyle{plain} %

\section{Introduction}

Transactional databases play a crucial role in today's digital world,
yet ensuring their correctness
is challenging due to bugs~\cite{cui2024understanding,jiang2023detecting,dou2023detecting},
failures~\cite{alquraan2018analysis,zheng2014torturing},
errors~\cite{dixit2021silent,wang2023understanding},
and security threats~\cite{tan20cobra,xia2022litmus,yang2020ledgerdb}.
Now, consider an ``oracle''
that serves as a decisive mechanism:
given a database execution trace,
it can either confirm correctness or reject it,
providing explanations for any incorrect behavior.
This oracle would be extremely valuable across various applications.
During testing, it verifies whether a database behaves as expected under
different workloads.
In auditing, it provides a thorough review of potential
errors and failures arising from underlying software and hardware issues.
For outsourced computing environments where trust cannot be assumed,
the oracle validates data integrity and execution against potential tampering.

In this work, we aim to build a practical approximate oracle. 
We introduce \sys, a checker framework that validates the correctness of transactional key-value store executions with sound and complete verification results.  
We focus on transactional key-value stores 
(which we refer to as ``databases'' throughout this paper for brevity)
as they are widely used in practice and serve as core building blocks for many complex systems, such as SQL databases~\cite{huang20tidb,cockroach_paper}
and file systems~\cite{juicefs}.

What distinguishes \sys from existing works is its focus on practicality. \sys makes significant advances in three key aspects: 

\begin{myenumerate2}
\item \sys makes no assumptions about the underlying database and
    supports arbitrary operations provided by modern transactional key-value stores.
    Previous works typically make some assumptions and restrict operation types,
        such as supporting only read and write operations. 
    In practice, however, all operations---including range queries and iterators---are widely used.

\item \sys does not require users to understand database internals
    (e.g., the concurrency control protocols in use)
    and enables checking with minimal logging.
    It uses a lightweight logging scheme called \emph{envelope tracing},
    which can be easily adopted without modifying application logic or the database---making it accessible to non-experts.

\item \sys features a modular pipeline that can be easily extended to support
    new correctness properties and configurations. Users can check new
        isolation levels and consistency properties by
        implementing modules that interact with \sys's intermediate representation (IR).
\end{myenumerate2} 
\sys is the first to support all three aspects above.
Previous black-box checkers such as Cobra~\cite{tan20cobra},
Viper~\cite{zhang2023viper}, and PolySI~\cite{huang2023efficient} cannot
support aspects (1) and (3)---they focus on one specific isolation level
and cannot handle a wide range of operations or new configurations.
Previous white-box checkers like Emme~\cite{clark2024validating} cannot
support aspects (2) and (3)---they require database expertise
and rely on parsing database internals using specialized tools (e.g., CDC tools~\cite[\S7.1]{clark2024validating}).
In fact, no system to date achieves aspect (3) and supports custom extensions.

\sys adopts a modular pipeline similar to a compiler's workflow:
(1) it logs inputs to and outputs from the database from the client's perspective, creating a trace that represents the database execution. %
(2) it parses the trace into a data structure called the \emph{\asg} (ASG).
The ASG contains the necessary information to reason about trace correctness. 
It extends a simple graph with additional metadata needed to check different correctness properties. 
(3) \sys lowers the ASG into low-level IR via semantic analysis. 
The IR layer introduces a key structure called \emph{\superpos}, which represents a set of potential scenarios that could explain the trace. 
We use \superpos to support arbitrary operation types and correctness properties. 
The semantic analysis also accepts user-defined hints to accommodate different setups and assumptions. %
For example, users can assert that all writes produce unique values, and \sys will incorporate this information in its analysis. 
(4) \sys applies optimizations and generates SMT constraints from the IR.
In particular, \sys introduces three optimizations: reachability pruning, topological prioritization, and unsat search.
(5) \sys solves the SMT constraints using a SAT solver and reports the results.

We have implemented \sys and evaluated it against state-of-the-art checkers. 
Its modular and generalized design allows \sys to verify all correctness properties and configurations supported by prior works. 
Functionality-wise, \sys is a superset of prior works. 
Performance-wise, \sys achieves comparable or better performance than state-of-the-art checkers. 
Notably, the core encodings of prior works can be implemented in \sys with just 
271--386
lines of code, %
demonstrating \sys's generalizability.

We evaluate \sys's practicality by exploring configurations not covered by prior works:
(1) implementing five previously unchecked isolation levels---repeatable read~\cite{berenson95critique,adya99weak},
read committed~\cite{adya99weak,berenson95critique},
PL-2+~\cite{adya99weak}, cursor stability~\cite{adya99weak}, and PL-FCV~\cite{adya99weak}
and two new encodings for serializability and snapshot isolation;
(2) testing a popular real-world application---a distributed file system JuiceFS~\cite{juicefs}---to audit its metadata management; 
(3) supporting a vendor-specific behavior of \ttt{SELECT FOR UPDATE} in MariaDB that breaks traditional transactions' visibility;
(4) confirming the previously theorized timestamp inversion pitfall~\cite{lu23ncc} in a real implementation. Meanwhile, it also discovers a new bug~\cite{newbug} in TiDB while cross-comparing behaviors across databases.

In addition to practicality, \sys offers further benefits as a general and extensible framework. 
First, it enables a fair comparison between different checkers. 
For example, while Viper and PolySI are both SI checkers, and previous results showed Viper to be slower than PolySI,  
implementing both in \sys reveals comparable performance, indicating that the reported performance gap stemmed from implementation differences rather than inherent checker differences. 
Another benefit is that \sys can share optimizations across different configurations. 
For instance, the topological prioritization is enabled by default for all checkers. 

\paragraph{Limitations.}
\sys is not without limitations. 
Supporting unmodified databases and arbitrary operation types means dealing with theoretically
NP-Complete problems~\cite{papadimitriou79serializability,biswas19complexity,bouajjani17verifying}. 
Like other black-box checkers, \sys cannot guarantee verification of all traces within bounded time. 
However, in practice, it remains useful as most traces are decidable, particularly with user-defined hints. 
Another limitation is that \sys does not support full SQL semantics, a limitation shared by all existing checkers. 
While \sys's design is, in principle, general enough to accommodate SQL semantics,
efficiently translating rich SQL semantics to SMT constraints
remains an open problem~\cite{pick2025checking}. %
\section{Setup, motivation, and applicability}
\label{s:s2}
\label{s:setup}

\subsection{Problem setup}

\begin{figure}[t]
\centering
\includegraphics[width=0.35\textwidth]{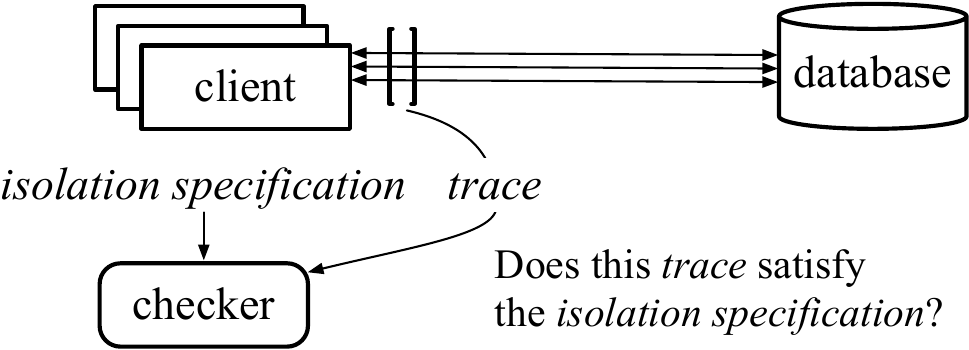}
\caption{Problem setup.}
\label{fig:setup}
\end{figure}

In this paper, we consider
the following setup.
Multiple clients simultaneously send requests to a database.
The database operates on the requests in a concurrent manner
and sends responses back to the clients.
These clients and the external world observe
the requests to and responses from the database,
which we refer to as a \emph{\trace}.
A checker then examines this \trace and determines whether the database
faithfully executes clients' requests.
Figure~\ref{fig:setup} depicts this setup.

We consider transactional operations:
all client operations are wrapped in transactions.
The operations that today's transactional key-value stores provide
are: puts, gets, deletions, range queries, and \iters.

The database claims to uphold some isolation specifications.
A canonical example is \ser, an \emph{isolation level}
defined by ANSI/ISO and widely adopted in the database community~\cite{adya99weak,crooks17seeing}.
These specifications are typically configured by the client,
establishing an agreement between the client and the database
on the expected execution semantics and transaction isolation guarantees.

This agreed-upon isolation specification is verified by a \checker.
The \checker evaluates the \trace %
against the specification
and returns a verdict: accept if the database
satisfies the isolation guarantees, or reject if it violates them.
Crucially, the checking must be sound and complete:
\begin{myitemize2}
    \item \emph{Soundness}: If the \checker accepts, the \trace satisfies the specification.
        (That is, if the \trace violates the isolation guarantees, the \checker will reject.)
    \item \emph{Completeness}: If the \trace satisfies the specification, the \checker will accept.
        (That is, if the \checker rejects, the \trace violates the isolation guarantees.)
\end{myitemize2}
There are checkers~\cite{li2023leopard,zellag12consistent,golab14client,wada11data,rahman12toward}
that sacrifice either completeness or soundness or both for performance,
which is not the target of this paper.
We focus on complete and sound database checking.

\subsection{Motivating examples}
\label{ss:examples}

Our goal is to make isolation checking practical.
Many challenges arise in practice.
For example,
advanced operators, such as range queries and iterators,
introduce complex semantics that are difficult to formalize and verify.
Vendor-specific isolation behaviors diverge from textbook definitions,
requiring a checker to adapt its reasoning to vendor-specific semantics.
Real-world workloads sometimes include duplicate values and other patterns that
violate common assumptions, yet these patterns are essential for exposing real-world bugs.
Modern systems also support isolation specifications from different semantic domains---for example,
combining isolation levels with read-your-own-write guarantees.
Together,
these challenges motivate a unified framework that is flexible and extensible
enough to support diverse operators, semantics, and isolation specifications.

Below we highlight four motivating examples that emphasize
the need and urgency of a unified checker framework.

\label{case:fs}
\heading{Case \#1: checking real-world workloads and applications.}
JuiceFS~\cite{juicefs}, a distributed file system, shows why a unified checking framework is needed.
It stores all metadata in a transactional key-value store,
where metadata inconsistencies can cause severe failures.
End-to-end auditing of metadata operations with a checker is therefore desirable,
as it provides strong correctness assurance.
However, real-world systems like JuiceFS rely on the full operator set,
including iterators, which no existing checker supports.
Existing theories cannot model an iterator's unknown outcomes.
When an item is not returned by an iterator,
the checker must distinguish whether it was never inserted, deleted, or
skipped by an incorrect iterator.
Without a way to express and resolve these ambiguities,
prior tools cannot verify systems like JuiceFS.

\label{case:forupdate}
\heading{Case \#2: supporting vendor-specific semantics.}
Vendor-specific concurrency behavior introduces further challenges for isolation checking.
For example, under the \emph{Repeatable Read} isolation level,
many databases implement \ttt{SELECT FOR UPDATE} as returning the most recent
version of a record rather than a transactionally consistent snapshot.
Vendors treat this as a feature, but it diverges from the textbook definition,
which requires all reads from a transaction to come from one consistent snapshot.
Consequently, existing checkers reject these valid executions.
An extensible framework can incorporate such semantics explicitly---for example,
by modeling operators that observe the most recent version---thereby enabling
correct reasoning across diverse database behaviors.

\label{case:uniqueval}
\heading{Case \#3: relaxing assumptions that may not hold in practice.}
Existing checkers make assumptions about workloads and applications.
A prominent example is that all values written to the same key are unique.
This assumption simplifies checking by avoiding ambiguity when multiple
transactions write identical values,
but real-world applications often produce duplicates,
and some correctness violations arise only in such cases.
In MariaDB, %
for example, a bug manifests
only when duplicate values are written~\cite{mariadbbug}.

Another assumption is the use of \emph{tombstones}:
instead of modeling physical deletion, checkers assume that the system inserts a logical marker.
Databases do not implement deletion this way.
These assumptions exist because they simplify checking;
relaxing them requires rethinking the core abstractions.
A practical checker must operate beyond the limited, distortion-prone representations
used in the simplified setup.

\label{case:}
\heading{Case \#4: checking combined isolation guarantees.}
In practice, systems expose combinations of isolation guarantees.
For example, databases that support both \ser and read committed
may differ in whether a transaction can read its own writes.
Such variations cannot be handled by simple patches,
since these behaviors must be encoded as part of the
isolation specification rather than bolted on afterward.
As a technicality,
serializability relies on \emph{anti-dependencies},
a concept that does not exist in read committed;
so when modeling read-your-own-write for read committed,
one cannot simply ``turn off'' anti-dependencies without redefining the model itself.
Supporting such diversity requires a framework with modular structure
and suitable abstractions so that new concurrency constraints
can be expressed and checked without redesigning the entire system.

\begin{figure*}[t]
\centering
\includegraphics[width=0.85\textwidth]{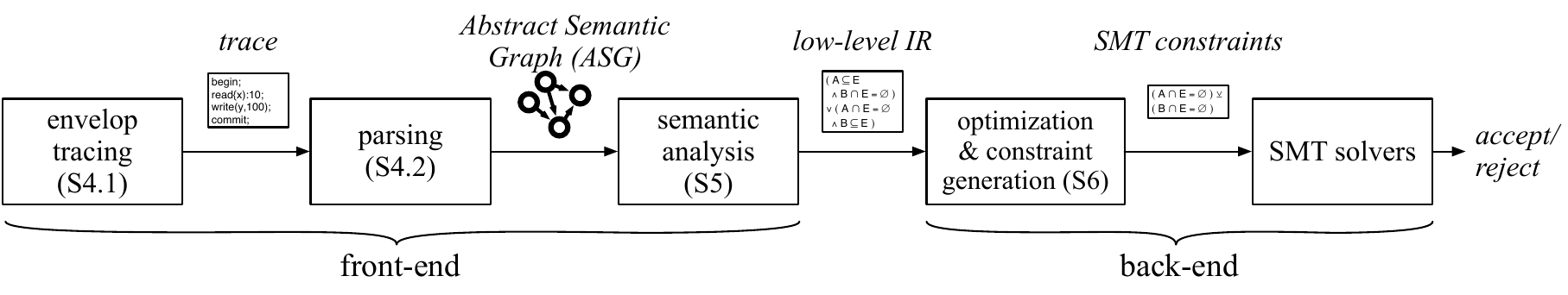}
\caption{\sys pipeline.}
\label{fig:pipeline}
\end{figure*}

\subsection{Practical applicability: illustrative scenarios}
\label{ss:applicability}

We view the isolation-checking framework as infrastructure for validating
database concurrency correctness.
It supports a wide range of use cases.
Below, we present three scenarios that illustrate how it can be applied in practice.

\heading{Scenario \#1: serving as a test oracle for concurrency testing.}
Concurrency testing exercises systems under diverse interleavings.
It requires a test oracle to determine whether an execution violates
the intended isolation guarantees:
for example, returning a value that should be impossible under given isolation level.
Existing frameworks such as Jepsen~\cite{jepsen-site}
rely on custom, condition-specific checkers that often sacrifice completeness or soundness.
A unified checking framework fills this gap by serving as a precise oracle:
it ingests execution traces,
evaluates them against the specified concurrency semantics (e.g., isolation levels),
and reports whether the behavior is correct.
Such principled validation is not possible today without comprehensive support for
real-world workloads and applications.

\heading{Scenario \#2: acting as an auditor for data integrity.}
Failures and errors are inevitable in both software and hardware.
On the one hand,
database users can legitimately wonder whether their data
maintains integrity throughout various failures and subsequent recoveries,
while continuously executing new transactions.
On the other hand, there is silent data corruption due to hardware errors.
Recent research~\cite{dixit2021silent,wang2023understanding} shows that
even CPUs sometimes silently produce incorrect results.
A \checker can play a critical role in detecting such corruptions,
as it checks final outcomes without relying on internal information.
\heading{Scenario \#3: verifying outsourced databases.}
Outsourced services are increasingly popular.
An example is cloud databases.
The database owners, who are also cloud users,
may have concerns about whether their databases perform
correctly.
This is especially true for correctness-critical applications
such as financial services and healthcare.
The database owner can use a \checker to
ensure the integrity of their outsourced databases
without trusting the cloud.

\heading{More potential applications.}
More broadly, a \checker enables unexplored applications:
auditing internal databases against external attacks for security,
proving database integrity for forensics,
and distinguishing suitable isolation levels for performance tuning.
These highlight a chicken-and-egg problem---without a unified, extensible framework,
researchers are unable to build checkers for niche use cases,
but without demonstrated use cases,
no one builds the extensible framework.
We hope our isolation-checking framework, \sys, breaks this cycle and opens new research directions.

\section{System overview}

This section gives an overview of \sys.
\sys accepts a trace and checks if the trace satisfies
expected isolation levels and other properties.
\sys provides native support for major isolation levels,
including strict serializability, serializability, snapshot isolation,
and read committed. %
In addition, \sys exposes an intermediate representation (IR) that enables users
to extend \trace semantics and define custom isolation properties
by implementing modular extensions.
Figure~\ref{fig:pipeline} depicts the workflow of \sys.

\heading{System overview.}
At a high level,
\sys translates a \trace, under a specified \iso,
into a set of logical formulas that are machine checkable.
This process mirrors traditional program compilation:
the \trace corresponds to the program,
and different isolation levels correspond to different programming languages,
each with its own semantics.
\sys introduces an abstract semantic data structure
and an intermediate representation (IR)
that capture the common structure of isolation checking.
The IR is then lowered into logical formulas---SMT constraints---that are
solved by a specialized SMT solver.

\sys's generalizability comes from its separation of front-end and back-end.
To support a new isolation level (e.g., PL-2+~\cite[\S4.1]{adya99weak})
or a variant of an existing one (e.g., snapshot isolation variants~\cite[Fig4]{crooks17seeing}),
people need to only modify the front-end\footnote{For some isolation levels, only back-end needs modifying.}---analogous to adding a new
programming language or dialect.
Meanwhile,
if new optimization strategies or specialized SMT techniques become available,
they can be integrated into the back-end and automatically benefit all
isolation-checking tasks.

This separation is enabled by \sys's IR.
At its core is a unified abstraction: \emph{\superposs} (\S\ref{ss:superpos}),
a concept that captures unknown dependencies in the \trace.
\CF{\superposs} encode every feasible way the \trace could have been generated by a database
that correctly enforces the specified isolation level.
By solving the resulting constraints, the system produces either a witness demonstrating correctness or a counterexample revealing a violation.

\heading{Pipeline design of \sys.}
\sys employs a pipeline that systematically transforms
traces into constraints.
This modular design enables flexible isolation checking and adapts to diverse trace details and isolation specifications.

\emph{Phase 1: tracing.}
\sys introduces a simple and intuitive tracing mechanism called \emph{envelope tracing} (\S\ref{ss:envelopetracing}).
By instrumenting \emph{only} the code immediately before and after each database
invocation, the tracer collects all information required for checking without intruding on application logic.
The interface is flexible: users can supply
additional hints that capture domain-specific knowledge relevant to checking.
These hints can later be processed through \emph{user-defined modules} (\S\ref{ss:usermodules})
and used to accelerate the checking pipeline.

\emph{Phase 2: parsing.}
In the parsing phase, \sys constructs an \emph{Abstract Semantic Graph (ASG)}
that captures the structural information embedded in the \trace.
The ASG consists of three key components: a graph representation, a \readinfo, and
a \writeinfo.
This phase effectively converts raw trace strings into a structured representation
suitable for further analysis.

\emph{Phase 3: semantic analysis.}
The semantic analysis phase transforms the ASG into a low-level
intermediate representation (IR) that we designed.
This IR includes simple graphs,
first-order logic expressions that describe
relationships between edges,
and \superposs.
\CF{\superposs} represent %
``known unknowns''---ambiguities in the \trace
where certain relationships must exist but cannot be precisely determined from
the available information.
For example, \superposs can capture the
uncertainty in the ordering of concurrent write operations:
they are conflicting and therefore must have an order, but without internal scheduling information
we cannot determine their order.
Meanwhile, if a \trace
provides complete information without ambiguities,
no \superposs will be generated. %
User-defined modules apply at this phase.

\emph{Phase 4: optimization and constraint generation.}
This phase encodes the low-level IR into a set of constraints suitable for SMT solving.
While most encodings follow straightforward translations,
\superposs require special handling and are encoded as mutually exclusive options.
Optimization passes are applied during this phase to improve solver performance.
These optimizations traverse the IR and apply equivalent
transformations that preserve correctness while producing constraints
that can be solved more efficiently.

\emph{Phase 5: solving.}
In the final phase, \sys leverages SMT solvers to determine
whether the \trace satisfies the specified isolation level.
By default, \sys uses MonoSAT, an SMT solver with native support for graph primitives,
making it particularly effective for the graph-based constraints generated in
the previous phase.
For broader compatibility, \sys can also compile its
constraints to classic SMT clauses suitable for other solvers such as Z3.
The solver produces either a \ttt{sat} result, confirming the trace's compliance
with the specified isolation level, or an \ttt{unsat} result, accompanied by a
counterexample illustrating the violation.
\section{Tracing and parsing}

\subsection{Client interface and envelope tracing}
\label{ss:envelopetracing}

In practice, clients interact with databases through various interfaces. They
may use standard libraries such as JDBC, or language-specific APIs provided by
the database. While these interfaces may differ in
aspects like function names and argument conventions,
modern transactional key-value stores expose the same
core functionality.
Here are all operations that today's production systems provide:
\begin{itemize}
    \item \emph{Transaction control}: \ttt{begin}, \ttt{commit}, \ttt{abort}
    \item \emph{Point queries}: \ttt{put}, \ttt{get}, \ttt{delete}
    \item \emph{Predicate-based queries}: \ttt{scan}, \ttt{\iter}
\end{itemize}

Most of these operations are self-explanatory.
The range query \ttt{scan} retrieves all items within
a specified key range \ttt{[start, end)}.
The \ttt{\iter} enables precise control over range scans,
supporting use cases like top-K
queries by allowing clients to fetch up to K items incrementally.
For example, an iterator can be initialized as \texttt{it$\gets$Iter(key0, key9)},
and the client then uses \texttt{it.Next()} to retrieve the next item iteratively.

Although widely used,
certain operations---such as \ttt{\iter} and
combinations like \ttt{delete} with \ttt{scan}---are not
supported by existing checkers.
This limitation is due to several challenges, including verifying that returned
items fall within the top-K, and that non-returned items either lie outside the
specified range or do not exist---conditions that are often partially unknown in the \trace.
We will elaborate on these challenges later (\S\ref{ss:challenge}).

\heading{Envelope tracing.}
Logging the necessary information in the \trace is essential for correctness checking.
However, instrumenting databases is difficult---if not impossible---especially
for modern distributed databases that are highly optimized and continuously evolving.
Parsing database metadata also requires significant expertise.
For example, Emme~\cite{clark2024validating} reconstructs execution traces by
parsing change data capture (CDC) logs: it uses the \texttt{CHANGEFEED}
mechanism for CockroachDB, the Debezium CDC tool for PostgreSQL, and heap
scanning for TiDB~\cite[\S7.1]{clark2024validating}.

To enable practical logging,
we advocate \emph{envelope tracing}---only logging the inputs and outputs of database operations:
\begin{lstlisting}[language=Python, 
                   basicstyle=\ttfamily\scriptsize,
                   keywordstyle=\color{blue}\bfseries,
                   commentstyle=\color{green!60!black}\bfseries,
                   stringstyle=\color{orange},
                   keywords={log_input, log_output},
                   showstringspaces=false]
  # "ctx" is a dictionary, containing...
  # ...context-specified information (e.g., hints)
  log_input(op, args, ctx) # log inputs to database
  ret = op(args)           # issue op to database
  log_output(ret)          # log output from database
\end{lstlisting}
Envelope tracing does not require modifications
to either the application logic or the databases.
Developers only need to instrument the interfaces.
For \sys, we create a set of wrappers for different databases
and their APIs. %

Envelope tracing captures only the minimal information necessary for checking.
As a result, the checking problem becomes challenging.
This trade-off is expected: more detailed traces simplify checking and improve checking performance,
but at the cost of more intrusive logging, tighter coupling to specific
applications, or reliance on database internals.
Existing checkers typically choose a fixed point along this trade-off curve,
while \sys supports the full spectrum.

\sys adopts a modular design. By default, it accepts the minimal trace from envelope tracing.
Yet, when additional information---for example, the information extracted from CDC logs---is available,
developers can write a module for the semantic analysis phase (\S\ref{s:analysis})
and \sys can then leverage it to improve checking efficiency.
Meanwhile, \sys provides common modules to manage trade-offs that have been explored by existing checkers.

\subsection{Parsing: building the ASG}

With a given \trace,
\sys constructs
an \emph{\asg (ASG)}: an intermediate structure that enables further lowering.
A crucial design goal is to ensure that ASG is both
general---capturing all necessary information across different isolation
levels---and efficient---allowing quick retrieval of relevant information
during the generation of low-level IR.

\label{ss:ir}
\label{ss:asg}

\heading{Observations.}
To design ASG,
we have two key observations.
First, isolation levels can be effectively expressed through graph
representations as highlighted
by prior work~\cite{bernstein79formal,papadimitriou79serializability,adya99weak,adya00generalized,weikum01transactional,zhang2024simplifying},
suggesting that graphs serve as a natural intermediate representation.
This graphical formulation allows us to capture the complex relationships
between transactions in an intuitive manner.
Second, we have found that all dependencies in different isolation levels
can be expressed as a set of conditioned read/write primitives.
At their core, the basic units of transaction operations are simply reads and
writes, but these are grouped by a potentially complex set of
conditions that determine their valid execution given different
traced information.

\begin{figure}[t]
\centering
\includegraphics[width=0.35\textwidth]{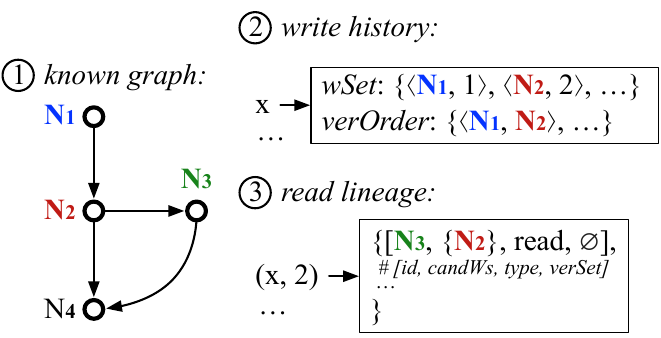}
\caption{An \asg example for serializability.
Nodes ``$N_i$'' represent transactions; ``x'' denotes a key;
and numbers such as ``1'' and ``2'' represent values.\\
In the \writeinfo, key x has a write set ``\t{wSet}'' consisting of node $N_1$ writing x=1
and node $N_2$ writing x=2. The version order ``\t{verOrder}'' specifies
that $N_1$ happens before $N_2$.\\
In the \readinfo, reading $x\to 2$ happens at least once for the node $N_3$
that reads from a set of candidate writers ``\t{candWs}'', which in this case contains only $N_2$.
This is an operation of type ``read'' with empty version set ``\t{verSet}''.\\
For different levels of trace detail,
some information (e.g., ``\t{verOrder}'' in \writeinfo) may be missing.
}
\label{fig:asg}
    \vspace{-2ex}
\end{figure}
 
ASG represents a new design that generalizes to different checking problems,
in contrast to prior checkers, which target a specific isolation level and therefore do not require generalization.
In particular, unlike previous fixed graph structures (e.g., serialization graphs~\cite{adya99weak}, polygraphs~\cite{tan20cobra}, and BC-graphs~\cite{zhang2024simplifying}),
ASG's flexible design adapts nodes and edges based on the target isolation level.
ASG also generalizes across all configurations, supporting white-, gray-, and
black-box traces; range and point queries; logical or physical deletions; and
unique or duplicated values.

\heading{ASG design.}
Based on these observations, we define our \asg (ASG) as
consisting of three essential components:
\circledone, a \emph{known graph};
\circledtwo, a \emph{\writeinfo};
and \circledthree, a \emph{\readinfo}.
Figure~\ref{fig:asg} depicts an example of ASG.
The known graph captures known relationships between
transactions, while the \writeinfo and \readinfo
store detailed information about write/read primitives.
Additional technical details of the ASG data structure and its construction
appear in Appendix A.
\section{Semantic analysis with \superposs}
\label{s:analysis}

The semantic analysis phase lowers the ASG to \sys's low-level IR,
which consists of simple directed graphs,
first-order logic expressions,
and \superposs---a new abstraction.
\sys introduces \superposs to address a class of challenges
in verifying transaction isolation in practice.

\begin{figure*}[t]
\vspace{-1ex}
\centering
\includegraphics[width=0.9\textwidth]{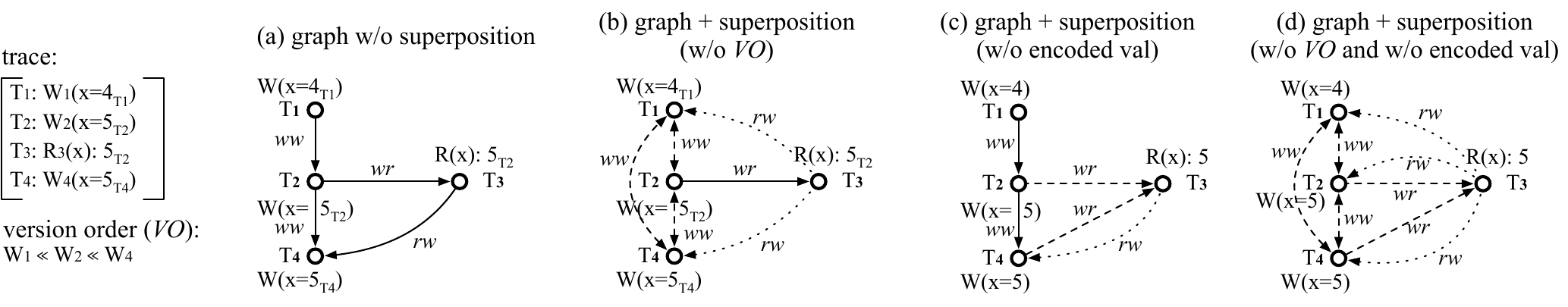}
\caption{\small
    An example demonstrating \unknowns and graphs with \superpos.
\textbf{On the left} is a trace with four transactions (begins and commits are omitted).
\textbf{On the right} are four graphs with different assumptions for checking \ser;
they all have transactions as nodes with solid edges representing known dependencies.
(a) is a graph with known version order and encoded values.
Graph (b), (c) and (d) incorporate \superpos:
dashed edges indicate possible dependencies from \superpos,
while dotted edges represent inferred \rwrels.
Specifically:
(b) excludes the version order but has the encoded values;
(c) includes the version order but omits the encoded values; and
(d) excludes both the version order and the encoded values.
}
\label{fig:example}
    \vspace{-1ex}
\end{figure*}

\subsection{Challenges in practice}
\label{ss:challenges}
\label{ss:challenge}
\label{ss:unknown}
\label{ss:unknowns}

The main challenge for practical isolation checking
is handling the ``known unknowns''---dependencies between transactions and operations
that are known to exist but cannot be fully identified due to missing information.
Here are two concrete examples:

\begin{myenumerate2}
    \item \emph{Phantom ambiguity}:
        consider a range query \ttt{scan(x,z)} where \ttt{y} $\in$  \ttt{[x,z)} is not returned.
        It is \textbf{unknown} whether \ttt{y} was never inserted or was
        inserted and subsequently deleted---especially when both \ttt{put(y,}$\dots$\ttt{)} and \ttt{del(y)} exist in the trace.
        This ambiguity becomes more challenging when there are multiple \ttt{put} and \ttt{del} on \ttt{y}.

    \item \emph{Read lineage uncertainty}:
       in real-world workloads, it is common for transactions to write duplicate values.
       For example, a trace may contain two committed transactions that each perform \ttt{put(x,9)},
       followed by a subsequent transaction that reads $x\to9$.
       In this case, it is \textbf{unknown} which transaction the read actually depends on.
\end{myenumerate2}%
These are just two examples; other unknown cases exist, some of which are
intertwined, for example, a trace
with range queries, \truedeletions, and duplicate writes.

%
%

\subsection{Abstracting unknowns using \superpos}
\label{ss:superpos}

We introduce \emph{\superpos}, a unified representation
that abstracts \unknowns.
We borrow the term from quantum mechanics;
\superpos indicates
a situation where two or more states are added
together (superposed).
However, when measured, the \superpos commits to a single state.

\Superpos is defined as follows:
\[
  \langle \posb_1,\ \posb_2,\ \cdots,\ \posb_n \rangle
\]
It is a tuple of $n$ \emph{\posbs}.
Each \posb indicates one way to explain the \obss.
In reality,
one and only one \posb is true,
but due to \unknowns, it is unclear which one is the ground truth.

For example, if a black-box database is serializable,
two concurrent transactions (say $T_i$ and $T_j$)
writing to the same key must be ordered.
However, due to the black-box setup, the
order of the two writes is unknown, leading to
a \superpos:%
\[
  \langle T_i \wwarrowx T_j,\ T_j \wwarrowx T_i \rangle,
\]%
reading as, either $T_i$ commits before $T_j$
or vice versa.

\heading{A toy example.}
We use a toy example to illustrate
why \unknowns are challenging %
and how \superpos abstracts \unknowns.
This example considers checking serializability for a \trace that
has four concurrent transactions, $T_1$ to $T_4$;
three write key $x$
and the other reads key $x$.
The values are encoded to include the write transaction id,
indicated by the subscript on values (e.g., $5_{T1}$).
Figure~\ref{fig:example} shows the example.

We start with no unknowns,
depicted by the leftmost graph in Figure~\ref{fig:example}.
The graph shows that the \trace is serializable,
because there exists a serial execution
of the four transactions: $T_1 \to T_2 \to T_3 \to T_4$.
The dependencies between writes ($T_1 \wwarrowx T_2$ and $T_2 \wwarrowx T_4$)
come from \vorder, which requires the white-box setup.
The dependency between writes and reads ($T_2 \wrarrowx T_3$)
is due to encoded values---$T_3$ reads the value $5_{T2}$ written by $T_2$.
Finally, $T_3$ must be executed before $T_4$
because otherwise the $T_3$ should have read $T_4$ instead of $T_2$;
this is \rwrels, which are considered
in \ser and snapshot isolation.

Next, we remove the white-box setup.
This means that internal scheduling information---\vorder---is unknown.
As a consequence, the dependencies between concurrent writes are unknown.
In our example, this means we know that $T_1$, $T_2$, and $T_4$ should have
a total order of writing to $x$, but the exact order is unknown.
Thus, pairwise \posbs are
(wrapped in \superpos):
$\langle T_1 \to T_2,\ T_2 \to T_1 \rangle$,
$\langle T_1 \to T_4,\ T_4 \to T_1 \rangle$,
$\langle T_2 \to T_4,\ T_4 \to T_2 \rangle$.
These are the double-arrowed dashed-edges in the second graph of Figure~\ref{fig:example}.
Also, as an implication of these \superpos,
there are two possible \rwrels, indicated by dotted edges.

If we restore the white-box setup but remove the encoded values,
a reading of value $5$ will not be enough to determine
whether the transaction reads from $T_2$ or $T_4$, because both write $x=5$.
The \superpos will be $\langle T_2 \to T_3,\ T_4 \to T_3 \rangle$,
indicated by dashed edges in the third graph of Figure~\ref{fig:example}.
The dotted edge again indicates a possible \rwrel.

Finally, consider the setup of a black-box database without encoded values:
we have all \superposs mentioned above.
Notice that \superposs compound:
there is a dotted edge, $T_3 \to T_2$, that does not appear
in any previous setups.
This edge is possible only when (1) $T_3$ reads from $T_4$
and (2) $T_4$ writes $x$ before $T_2$.
According to the ground truth,
neither condition-(1) nor condition-(2) is true;
however, they are possible given multiple types of unknowns.

%
\begin{figure}
\lstdefinelanguage{PythonHighlighted}{
  keywords={def, return, for, in, if, else, global},
  basicstyle=\ttfamily\scriptsize,
  keywordstyle=\color{blue}\bfseries,
  morekeywords=[2]{ASG, genIR, genReadDep, genSessionOrder, genSessionDep},
  keywordstyle=[2]{\color{purple}\bfseries},
  sensitive=true,
  comment=[l]{\#},
  commentstyle=\color{green!60!black},
  stringstyle=\color{red},
  morestring=[b]',
  morestring=[b]"
}

\begin{lstlisting}[language=PythonHighlighted,
                   numbers=left,
                   frame=lines,
                   framesep=3pt,
                   %
                   ]
global ASG # read only

def genIR():
  g,supp,exp = ASG.g,{},{}  # init low-level IR
  g,supp,exp = genReadDep(g, supp, exp)
  ... # the other dependency passes

  # user-defined modules
  userModules = [genSessionOrder, ...]
  # the "ctx" is from envelope tracing interface
  for f in userModules:
    g,supp,exp = f(g, supp, exp, ctx)

  return g,supp,exp

# an example of dependency pass
def genReadDep(g, supp, exp):
  lineage = ASG.readLineage
  for (k,v) in lineage:
    for entry in lineage[(k,v)]:
      if entry.candW.size() == 1:
        # if write is unique, add a known edge
        g.Edges.add(...)
      else:
        # if multiple possibilities, add a superposition
        supp.add(...)
  return g, supp, exp

# an example of user-defined module
def genSessionDep(g, supp, exp, ctx):
  for (id1, id2) in ctx['session']: # session order
    g.Edges.add(id1,id2)            # enforce this order
  return g, supp, exp
\end{lstlisting}
\caption{Transforming ASG to low-level IR. This describes the logical transformation
flow. In practice, the implementation combines common dependency passes to
improve efficiency.}
\label{fig:algo:analysis}
\end{figure}
 
\subsection{Semantic analysis: generating low-level IR}
\label{ss:encodesuperpos}

\headingzero{Intermediate representation (IR).}
In the semantic analysis phase, \sys transforms the ASG into low-level IR.
Figure~\ref{fig:algo:analysis} depicts this transformation.
\sys's low-level IR consists of three components:
a simple graph (\ttt{g} in Figure~\ref{fig:algo:analysis}),
logical expressions (\ttt{exp}) across edges---treated as boolean variables
where true indicates existence and false indicates absence---and \superposs
(\ttt{supp}) that capture unknowns in the trace.
The formal definition of the IR is in Appendix A.

At a high level, \sys processes the ASG through multiple passes,
generating each required dependency type one at a time.
Each dependency-generating pass
analyzes the ASG to determine whether a
dependency exists between two nodes.
Crucially, isolation levels define the conditions under which dependencies should exist,
but the trace might not offer enough information to definitively establish these conditions.
For example, when generating read dependencies in \ttt{genReadDep(...)},
if there is only one possible write that could have produced a
value, we add a known edge to the graph (line 22, Figure~\ref{fig:algo:analysis}).
However, when multiple possibilities exist,
we create a \superpos to represent this uncertainty (line 25, Figure~\ref{fig:algo:analysis}).
These superpositions, along with logical expressions over edges, form the core of our
low-level IR's ability to represent complex dependencies and unknowns.

\label{ss:usermodules}
\heading{User-defined modules.}
\sys's modular design enables extension through user-defined modules.
Users will write a module function (described in the figure below)
whose inputs are the low-level IR after common passes
and the hints logged in envelope tracing.
\begin{figure}[h]
\lstdefinelanguage{PythonHighlighted}{
  keywords={def, return, for, in, if, else, global},
  basicstyle=\ttfamily\scriptsize,
  keywordstyle=\color{blue}\bfseries,
  morekeywords=[2]{ASG, f, registerUserModule},
  keywordstyle=[2]{\color{purple}\bfseries},
  sensitive=true,
  comment=[l]{\#},
  commentstyle=\color{green!60!black},
  stringstyle=\color{red},
  morestring=[b]',
  morestring=[b]"
}
\vspace{-2ex}
\begin{lstlisting}[language=PythonHighlighted,
                   %
                   frame=lines,
                   framesep=3pt,
                   %
                   ]
global ASG # read only

def f(g, supp, exp, hints):
  ... # user can access ASG and update g, supp, exp
  return g, supp, exp

registerUserModule(f)  # add f to the list userModules
\end{lstlisting}
\vspace{-2ex}
\end{figure}
User modules also have access to the read-only ASG.
Within the module, users can manipulate the low-level IR to describe their own semantics.
Figure~\ref{fig:algo:analysis} gives a simple module example
supporting session ordering dependencies, \ttt{genSessionDep(...)}.
This extensibility ensures that \sys can adapt to different isolation
levels and incorporate various sources of information while maintaining a
consistent overall approach.
\section{Optimization and constraint generation}

As a unified framework,
\sys enables sharing optimizations across different encodings and checkers.
Below are three optimizations that significantly accelerate solving.

\label{ss:hints}
\label{ss:optimizations}
\heading{\CF{\optone}.}
Each \superpos contains multiple possibilities,
and solving involves searching through these possibilities.
However, off-the-shelf solving
treats all \posbs equally, searching them in a random order.
\CF{\optone} addresses this inefficiency by prioritizing \posbs
that are more likely to be true within the \superpos.
In particular, we topologically sort the existing graph
and use the resulting node order as the priority.
For example, consider a \superpos $\langle T_i \to T_j,\ T_j \to T_i \rangle$.
If $T_i$ is topologically ordered before $T_j$, the \checker then
prioritizes exploring $T_i \to T_j$ over the other possibility.

\heading{\CF{\opttwo}.}
Node reachability serves as a powerful tool to eliminate \superposs~\cite[\S3.3]{tan20cobra}.
For example,
when checking serializability,
if a path $T_i \rightsquigarrow T_j$ exists in the graph,
the \superpos $\langle T_i \to T_j,\ T_j \to T_i \rangle$ can be immediately resolved,
as $T_j \to T_i$ would introduce a cycle,
a violation to most isolation levels.\CP{revisit Jian's comment in ltx}
As an optimization, we precompute the graph's transitive closure,
enabling fast reachability checks that systematically prune \superpos,
reducing unnecessary exploration in the solving phase.

%
%
%
%
%
%
%
%
%
%
%
%
%
%
%
%
%
%

%
\heading{Improving rejection performance with \unsatsearch.}
\label{ss:unsatsearch}
Proving unsatisfiability is challenging for today's \checkers.
For example, Viper significantly underperforms
with unsat \traces, occasionally even timing out~\cite[Figure 15]{zhang2023viper}.
The fundamental reason is that
rejecting a non-trivial unsat \trace requires
proving that no set of \posbs satisfies
\specs.

Though challenging,
proving unsatisfiability for real-world database executions
exhibits some characteristics.
We observe that the unsat \traces usually have a small ``unsat core''---a small set of
transactions that already violate the \specs.
This observation has also been confirmed by other researchers:
Cui et al.~\cite{cui2024understanding} studied 140 real-world transaction bugs
and discovered that
almost all (93.6\%) of these bugs
involved no more than three transactions, with all bugs triggered by at most
five transactions~\cite[Finding 4]{cui2024understanding}.
Our hypothesis is that violations are usually caused by corner cases plus high
concurrency. Thus, the erroneously behaved transactions
usually occur in a relatively short period of time.

Leveraging this observation, our idea for proving unsatisfiability efficiently
is to divide a \trace into
smaller sub-\traces and verify each sub-\trace individually.
This strategy improves the efficiency of rejections,
as the number of possibilities decreases exponentially with the reduction in the number of transactions.
Consequently, this approach requires exponentially fewer searches to prove unsatisfiability. We refer to this method as \emph{\unsatsearch}.

The guarantee of \unsatsearch is that if it rejects,
then the original \trace is unsat.
This is because the unsat core identified during the \unsatsearch
is also a valid unsat core for the original \trace.
However, failing to find an unsat core
does not necessarily imply that the original \trace satisfies the \specs,
as unsat cores may span across sub-\traces.
Here, our earlier observation becomes critical: if unsat cores are typically small,
the likelihood of all these cores being split across different sub-\traces is low.
This likelihood is further reduced if the \checker strategically segments
the \trace at points where there are few concurrent transactions,
indicating lower levels of concurrency and contention.
Our implementation %
uses random segmentation;
exploring more efficient \trace segmentation strategies is our near-future work.

\label{ss:encoding}
\label{ss:encodespecs}
\label{ss:isolationlevel}
\heading{Constraint generation.}
After optimizations,
\sys emits constraints based on the low-level IR.
It creates a Boolean variable for each edge:
\t{True} represents the existence of an edge; \t{False} means otherwise.
For edges in the IR's graph \ttt{g},
\sys assigns them to be \t{True}.
For logical expressions \ttt{exp} like \irimplies (\S\ref{ss:irimplies}),
\sys emits corresponding SMT formulas directly.
For each \superpos in the set \ttt{supp},
say $\langle e_1, e_2, \dots, e_n \rangle$,
\sys ensures that one and only one of the possible expressions is \t{True}, by
$\bigvee_{i=1}^{n} \left( e_i \land \bigwedge_{j \neq i} \neg e_j \right)$.
Finally, \sys adds isolation correctness constraints---typically graph
acyclicity---based on the isolation level definition being verified.

\section{Implementation}
\label{s:impl}

We implement \sys in Java, consisting of approximately 14K %
lines of code (LOC). This includes around 0.6K %
LOC of building the ASG, 1.7K %
LOC of converting ASG to low-level IR, 1.1K %
LOC of optimizations, 1.9K %
LOC of translating IR to SMT clauses, and other utilities.
\section{Experimental evaluation}
\label{s:eval}

In this section, we evaluate \sys along three dimensions:
(1) its applicability across different configurations
(i.e., isolation levels, benchmarks, and databases);
(2) its effectiveness in practical use cases; and
(3) its performance.

\heading{Setup.}
All experiments are run on a 12-core
AMD Ryzen 5900 machine,
with 64GB of memory and Ubuntu 20.04.

\subsection{\sys works across all existing configurations}

\headingzero{Configuration space.}
We evaluate \sys across all known configurations,
in terms of isolation levels,
benchmarks,
and databases.
The full detailed configurations are described in Appendix A.

\headingzero{Isolation levels.}
We experiment with six well known isolation levels:
strict \ser (\emph{SSER}),
\ser (\emph{SER}),
snapshot isolation (\emph{SI}),
repeatable read (\emph{RR}),
read committed (\emph{RC}),
and \emph{PL-2+} defined by Adya~\cite[\S4.1]{adya99weak}.

\headingzero{Benchmarks.}
We experiment with nine existing benchmarks from prior work~\cite{tan20cobra}:
RUBiS~\cite{rubis},
Twitter~\cite{twitter},
TPC-C~\cite{tpcc},
BlindW-C~\cite{cobra_verifier,cobra_bench} from Cobra~\cite{tan20cobra};
ListAppend-J from Jepsen;
\blindwt~\cite{zhang2015} (to distinguish from Cobra's BlindW-C benchmark),
Retwis~\cite{zhang2015} from Tapir~\cite{zhang2015},
BlindW-E from Port~\cite{lu2020port};
and the built-in filesystem performance testing benchmark from JuiceFS~\cite{juicefs}.

We also reimplement BlindW (to distinguish from BlindW-C and BlindW-E).
Further, we implement three additional benchmarks focusing
on range queries, iterators, and \truedeletions, without the \uval assumption.

\begin{myitemize2}
    \item \emph{RandomBench}:
    this benchmark includes all the operations that we support, 
    including reads, writes, range queries, true deletions, and iterators.
    No prior black-box \checkers support this benchmark.
    Written values are non-unique,
    and each transaction has 8 operations.

    \item \emph{RandomRange}:
    this benchmark evaluates range queries.
    Each transaction has 8 operations,
    with 20\% range queries
        and 80\% other operations (reads, writes, deletions, and read-modify-writes).

    \item \emph{RandomRMWRange}: this benchmark simulates Emme's experiments~\cite{clark2024validating}.
    Clients issue transactions with 8 operations, each being a read, a write or a range query.
    Any write is preceded by a read of the same object, forming a read-modify-write.
    Any range query is preceded by a global range query to recover the version set.
\end{myitemize2}

\headingzero{Databases.}
We experiment with nine databases:
PostgreSQL, YugaByteDB, CockroachDB, TiKV, Tapir~\cite{zhang2015},
\Port~\cite{lu2020port}, MariaDB~\cite{mariadb}, MySQL~\cite{mysql}, and
TiDB~\cite{tidb}.

\heading{Experimental results.}
We evaluate the generalizability of \sys
by applying it to check all mentioned isolation levels across
a diverse set of benchmarks and databases.
Figure~\ref{fig:check-realworld-workloads3} shows a subset of the results.
Note that we present only a single trace for each configuration for
demonstration purposes, not for performance evaluation.
RandomRange and RandomBench can only be checked by \sys since they have operations that existing checkers cannot handle.
We only present the result of \Port on a relatively small \trace.
Even though both PL-2+ and \Port are related to the notion of transactional causal consistency,
it is unclear whether \Port provides PL-2+, but its traces pass our checking.
The benchmarks are all run as-is, %
without any modifications.

\begin{figure}
\centering
\scriptsize
\begin{tabular}{llrlr}
\hline
\textbf{Iso.} & \textbf{Database} & \textbf{\#Txns} & \textbf{App} &\textbf{Runtime} \\
\hline
  \multirow{2}{*}{SSER} & \multirow{2}{*}{Tapir~\cite{zhang2015}} & 13.2K & BlindW-T & 1.14s \\
   &  & 9.1K & Retwis &  0.63s \\
\hline %
  \multirow{2}{*}{SER} & \multirow{2}{*}{CockroachDB 25.1.4~\cite{cockroach_paper}} & 10.5K & BlindW-C & 4.45s \\
  &  & 18.2K & TPC-C & 18.54s \\
\hline
  \multirow{2}{*}{SER} & \multirow{2}{*}{PostgreSQL 15.2~\cite{postgresql_website}} & 9.8K & RUBiS & 2.81s \\
  &  & 10.1K & Twitter & 2.43s \\
\hline %
  \multirow{2}{*}{SER} & \multirow{2}{*}{YugaByteDB 2.25.1.0~\cite{yuga_git}} & 10.4K & BlindW-C & 4.4s \\
  &  & 15.3K & TPC-C & 10.65s \\
\hline
  \multirow{2}{*}{SI} & \multirow{2}{*}{TiKV 7.6.0~\cite{tikv}} & 11.5K & JuiceFS 1.2.3 & 10.03s \\
  &  & 10K & RandomBench & 370.81s \\
\hline %
  \multirow{2}{*}{SI} & \multirow{2}{*}{YugaByteDB 2.25.1.0~\cite{yuga_git}} & 10.3K & RUBiS & 4.08s \\
  &  & 10.4K & Twitter & 4.42s \\
\hline %
  \multirow{1}{*}{SI} & \multirow{1}{*}{TiDB 8.5.4~\cite{tidb}} & 10K & RandomRange & 16.76s \\
\hline %
  \multirow{1}{*}{RR} & \multirow{1}{*}{MariaDB 12.1.2~\cite{mariadb}} & 10K & BlindW & 3.25s \\
\hline %
  \multirow{1}{*}{RR} & \multirow{1}{*}{MySQL 9.5.0~\cite{mysql}} & 10K & BlindW & 2.47s \\
\hline
  \multirow{1}{*}{PL-2+} & \multirow{1}{*}{Port~\cite{lu2020port}} & 0.8K$^\star$ %
               & BlindW-E & 0.2s  \\
\hline
  \multirow{2}{*}{RC} & \multirow{2}{*}{PostgreSQL 15.2~\cite{postgresql_website}} & 10.1K & BlindW-C & 2.99s \\
  &  & 14.6K & TPC-C & 5.94s \\
\hline
  \multirow{1}{*}{RC} & \multirow{1}{*}{MariaDB 12.1.2~\cite{mariadb}} & 10K & BlindW & 5.33s\\
\hline
  \multirow{1}{*}{RC} & \multirow{1}{*}{MySQL 9.5.0~\cite{mysql}} & 10K & BlindW & 5.63s\\
\hline
\end{tabular}
\caption{\Sys can check various combinations of isolation levels, databases, and workloads.\\
$^\star$ Checking PL-2+ is challenging, and \sys does not scale to 10K transactions as it does for other isolation levels.
}
\label{fig:check-realworld-workloads3}
\vspace{-2ex}
\end{figure}

\subsection{Case Study}

\headingzero{Case \#1: discovering a new bug in TiDB.}
While running benchmarks on TiDB 8.5.4, its latest version,
\sys detected an isolation-level violation in pessimistic mode under repeatable read.
The anomaly's effect is that a read following an \ttt{UPDATE} statement can not see this write
if the update value
matches that of another transaction.
We reported the issue~\cite{newbug}, and the TiDB developers acknowledged it as a moderate-severity bug.

\headingzero{Case \#2: catching bugs in PostgreSQL 12.3 and MariaDB 10.6.}
\sys can serve as a test oracle for concurrency testing.
We evaluate its ability to detect concurrency bugs by deploying two known-buggy
databases---PostgreSQL 12.3 and MariaDB 10.6---running our benchmarks,
and using \sys as the oracle.
\sys reliably catches bugs within tens of runs.
\sys also uncovers multiple anomalies. For PostgreSQL, it identifies the
expected G2-item anomaly and additionally detects a general G2 cycle involving
range queries under serializability, which existing checkers cannot capture.
For MariaDB, \sys rediscovers that a read following an update does not observe the update but
instead reads from the transaction's initial snapshot under repeatable read.
This anomaly manifests only when non-unique values are used, a case again unsupported by existing checkers.

\headingzero{Case \#3: auditing file system metadata operations.}
JuiceFS uses TiKV as its metadata storage
and thus benefits from the correctness guarantees provided by ACID transactions.
The metadata integrity is critical to file system correctness,
and developers can use \sys to audit
if the underlying metadata storage actually provides the claimed guarantees.
Notably, JuiceFS uses all provided APIs, including \ttt{Get}/\ttt{Set}/\ttt{Iter}/\ttt{Delete};
no existing checkers can fully support them.
\sys can check JuiceFS's metadata interactions with TiKV.
One such result is reported
in the ``JuiceFS 1.2.3'' row of
Figure~\ref{fig:check-realworld-workloads3}.

\headingzero{Case \#4: customizing checkers for vendor-specific behavior.}
MariaDB's repeatable read mixes snapshot reads from MVCC with locking
reads and writes (e.g., \ttt{SELECT FOR UPDATE}) that observe the most recent
committed state.
As a result, a transaction does not operate under any standard
isolation level, and no existing checker can model this behavior.
With \sys, we customize the checker to support MariaDB's semantics.
We partition each physical transaction into multiple logical sub-transactions:
locking reads and writes start new sub-transactions with updated views,
while snapshot reads are assigned to the appropriate earlier sub-transaction.
This modification requires only 85 LOC changes in front-end,
where \sys parses and structures transactions accordingly.
With this customization, \sys correctly checks MariaDB's repeatable read
behavior and accepts traces consistent with MariaDB's intended isolation guarantees.
This allows \sys to bypass anomalies that are prohibited under the standard definition of repeatable read but permitted by MariaDB's design, thereby focusing only on genuine violations.
Note that this partitioning strategy is not a tight checker, as it intentionally relaxes certain intra-transaction constraints.

\headingzero{Case \#5: supporting mixed isolation guarantees.}
Beyond standard isolation levels, databases often provide additional guarantees
such as read-your-own-write (RYOW). In our experiments, we observe that
databases differ in whether and how they support RYOW within a transaction.
\sys accommodates these variations by implementing three RYOW policies within 38 LOC changes: must
read your own writes, must read externally, and may read from either.
These policies can be inferred automatically or specified by the user,
enabling flexible support for mixed isolation guarantees.

\headingzero{Case \#6: confirming a theoretical hypothesis: timestamp inversion pitfall.}
Haonan et al.~\cite{lu23ncc} identify a potential issue in many timestamp-based
transaction protocols, named \emph{timestamp inversion pitfall}.
They provided a counterexample in the appendix of their paper's extended version~\cite{lu23nccextended}.
Using \sys, we successfully confirm this phenomenon in practice:
we found a \trace from Tapir that is serializable
but not strictly serializable, demonstrating a real instance of the
timestamp inversion pitfall.

%
%
%
%
%
%
%
%

%
%
\begin{figure}
	\footnotesize
	\begin{tabular*}{\columnwidth}{@{\extracolsep{\fill}} llllc @{}}
		\toprule
		\textbf{Checker} &  \textbf{Gen. IR} & {\textbf{Pruning}} &  {\textbf{Gen. SMT}}   & {\textbf{Total LOC}}  \\
		\midrule
		Boomslang SSER                    & 55         & 10$^\dagger$  &  180$^+$        & 245   \\
		Boomslang SER                     & 51$^\star$ & 10$^\dagger$  &  180$^+$       & 241   \\
		Boomslang RR                     & 59 & 10$^\dagger$  &  180$^+$       & 249   \\
		Boomslang SI                      & 95         & 10$^\dagger$  &  180$^+$        & 285   \\
		Boomslang RC                      & 48         & 10$^\dagger$  &  180$^+$        & 238   \\
		Boomslang PL-2+                   & 51$^\star$ & 36$^\bullet$  &  178            & 265   \\
		Boomslang CS                      & 51$^\star$ & 36$^\bullet$  &  277            & 364   \\
		Boomslang PL-FCV                     & 51$^\star$ & 36$^\bullet$  &  226            & 313   \\
		\bottomrule
	\end{tabular*}
	\caption{Lines of code (LOC) for each checker component: IR generation (Gen. IR), pruning, and SMT encoding.\\
		$^\dagger,^\star,^\bullet,^+$ means this component is shared across checkers.
		For example, Boomslang SER/PL-2+/CS share the same component of converting ASG to low-level IRs.
	}
	\label{fig:loc-checkers2}
\end{figure}
 \headingzero{Case \#7: implementing new checkers made easy.}
\sys is an extensible framework.
To demonstrate this,
we implement eight different checkers
and measure their lines of code (LOC) for different phases.
Figure~\ref{fig:loc-checkers2} shows the results.

Among the eight checkers,
we implement five checkers Boomslang RR, RC, PL-2+, CS, and PL-FCV
that verify previously uncheckable isolation levels---repeatable read~\cite{berenson95critique,adya99weak}, 
read committed~\cite[Section 3.2.2]{adya99weak},
PL-2+~\cite[Section 4.1]{adya99weak}, 
cursor stability~\cite[Section 4.6]{adya99weak} 
and PL-FCV~\cite[Section 4.6]{adya99weak}.
Figure~\ref{fig:loc-checkers2} %
shows that all the checkers are implemented with $<$370 lines of code.
\sys makes module reuse easy:
three checkers---Boomslang SER, PL-2+, CS and PL-FCV---share the same conversion code from ASG to IR,
because their IR needs the same types of superpositions and first-order expressions.
Another five checkers (Boomslang SSER, SER, RR, SI, RC) share the same pruning strategies
because they all require no-cycle property in a graph with all dependencies.
\begin{figure}
	\footnotesize
	\begin{tabular*}{\columnwidth}{@{\extracolsep{\fill}} llllc @{}}
		\toprule
		\textbf{Checker}  & {\textbf{Original LOC}} & {\textbf{LOC in \sys}} \\
		\midrule
		Cobra~\cite{tan20cobra}            &  16.6K & 271 \\
		Viper~\cite{zhang2023viper}       &  2.9K & 281  \\
		PolySI~\cite{huang2023efficient}   &  3.4K & 386 \\
		\bottomrule
	\end{tabular*}
	\caption{Lines of code (LOC) for Cobra/Viper/PolySI's original system and their implementations in \sys.
	}
	\label{fig:loc-original-vs-ours}
\end{figure}
 \headingzero{Case \#8: replicating existing checkers.}
To further illustrate \sys's flexibility,
we reimplement three existing checkers: Cobra, Viper and PolySI.
Figure~\ref{fig:loc-original-vs-ours} shows the results.
We also include the LOC of their original implementations.
\sys can reduce the lines of code by about 9$\times$.

\headingzero{Case \#9: enabling fair comparison across checker encodings.}
\begin{figure}[t]
    \centering
    \includegraphics[width=0.43\textwidth]{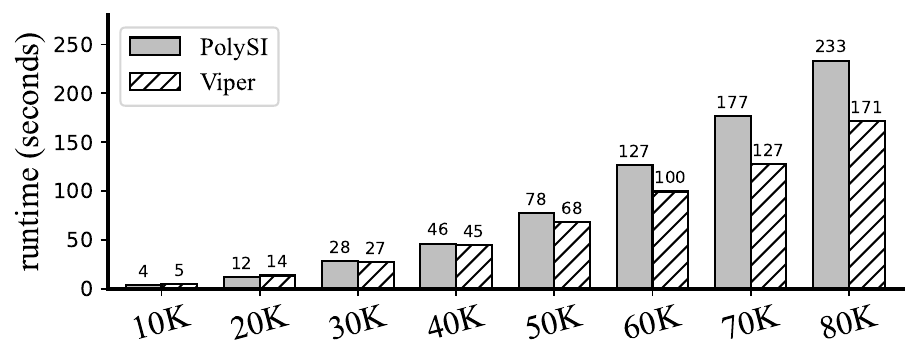}
    \caption{Checking SI: a fair comparison between PolySI's and Viper's encodings}
    \label{fig:faircomp2}
\end{figure}
 Multiple encodings have been proposed for checking the same isolation level,
such as snapshot isolation.
We reimplement these encodings in \sys to enable a fair performance comparison,
removing differences due to implementation-specific factors.
Figure~\ref{fig:faircomp2} compares two encodings---originally from Viper and
PolySI---on the same traces: the BlindW-C benchmark
with from 10K to 80K transactions.
Viper's original implementation is slower than PolySI's due to its Python
implementation and PolySI's specialized optimizations.
However, when comparing only the encodings,
Viper's encoding slightly outperforms PolySI's---an insight made
possible only by a unified framework like \sys.

\subsection{Performance evaluation}
\label{ss:baseline}

\headingzero{Baselines.}
We compare \sys with five state-of-the-art \checkers:
Cobra~\cite{cobra_verifier}, Viper~\cite{zhang2023viper,viper-repo}, PolySI~\cite{huang2023efficient,polysi-repo},
Elle~\cite{kingsbury20elle},
and Emme~\cite{clark2024validating}.
They are the most closely related work.
The first three work in a black-box setup and assume \uval.
Cobra verifies \ser;
Viper verifies \si and supports range queries on keys with \tombstone; %
PolySI verifies \si.
Elle is a grey-box checker that can check multiple isolation levels, but may miss anomalies.
It relies on the ListAppend benchmark to recover the version order to achieve soundness.
The last one, Emme, is a white-box checker.
With Emme's original implementation unavailable, we reimplement Emme in our framework \sys.

We compare \sys's performance with baselines
using CobraBench~\cite{cobra_bench}.
Benchmarks are executed on PostgreSQL with session orders, as required by the baselines.

\headingzero{Checking \ser: \sys versus Cobra.}
\begin{figure}[t]
    \vspace{-1ex}
    \centering
    \includegraphics[width=0.4\textwidth]{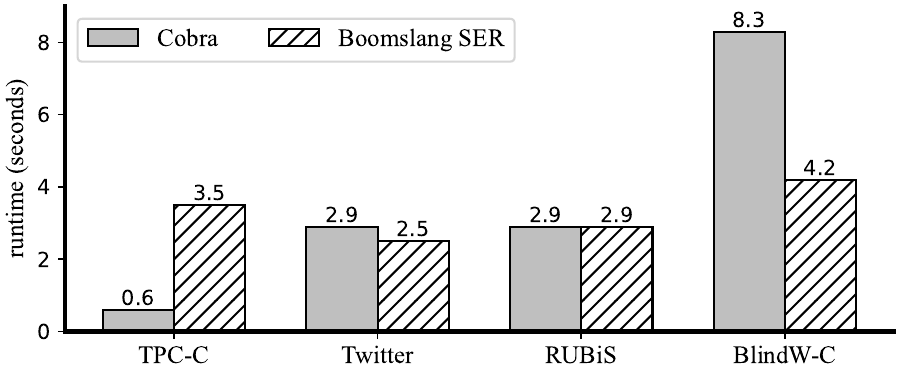}
    \caption{Checking serializability: \sys and Cobra.
    }
    \label{fig:ser-baselines}
\end{figure}
 We compare \sys with Cobra on the problem of checking serializability.
We run CobraBench with 10K transactions for each benchmark.
Cobra relies on GPU acceleration, and thus its scalability is fundamentally constrained by GPU memory.
On our 3060 Ti GPU, Cobra cannot handle 20K transactions and hence 10K transactions are used for the comparison.
Figure~\ref{fig:ser-baselines} shows the results.
Compared with Cobra, \sys achieves comparable or better results, except for TPC-C.
\Sys has additional \optone that Cobra does not have, %
which can guide the process of assigning values to edge variables, and hence \sys is faster than Cobra on
all the \traces that have \superposs.
Meanwhile, TPC-C does not have \superposs, and hence all the edges are known.
In such cases, \sys spends extra effort on bookkeeping edges
due to generality, whereas Cobra does not.

\headingzero{Checking SI: \sys versus Viper and PolySI.}
\begin{figure}[t]
    \centering
    \includegraphics[width=0.45\textwidth]{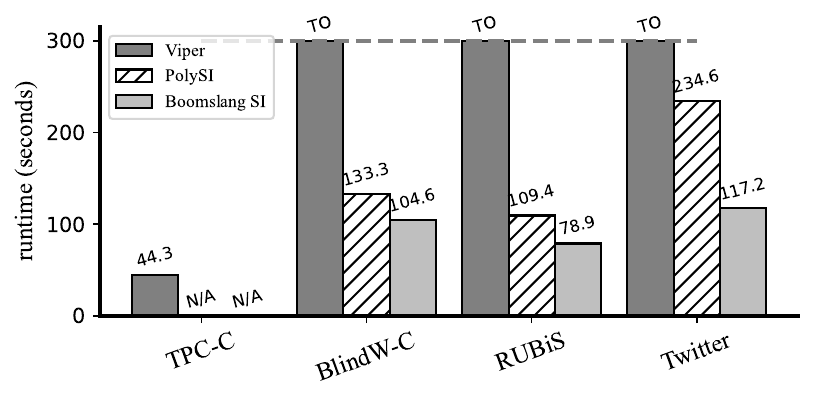}
    \caption{Checking SI: \sys, Viper, and PolySI.
    }
    \label{fig:si-baselines-50k}
    \vspace{-2ex}
\end{figure}
 Figure~\ref{fig:si-baselines-50k} compares the runtime performance of \Sys against Viper and PolySI across four workloads with 50K transactions. 
The results reveal distinct scalability bottlenecks in existing tools: Viper consistently hits the 600-second execution timeout (TO) on BlindW-C, RUBiS, and Twitter, indicating Viper fails to scale computationally for large workloads. In contrast, \sys demonstrates superior efficiency, successfully verifying these complex traces well within the time limit.
Furthermore, \sys consistently outperforms PolySI in terms of speed.
Notably, on the compute-intensive Twitter workload, \sys achieves a 2x speedup compared to PolySI (117.2s vs. 234.6s) and maintains a significant lead on RUBiS (78.9s vs. 109.4s). 
Both \sys and PolySI OOM on the TPC-C benchmark but handle the other three, which shows that \sys incurs comparable memory overhead as PolySI across the four workloads.
Interestingly, only Viper can check TPC-C, which is probably due to its unique heuristic pruning.
\headingzero{Checking \ser: \sys versus Elle.}
\begin{figure}[t]
    \centering
    \includegraphics[width=0.4\textwidth]{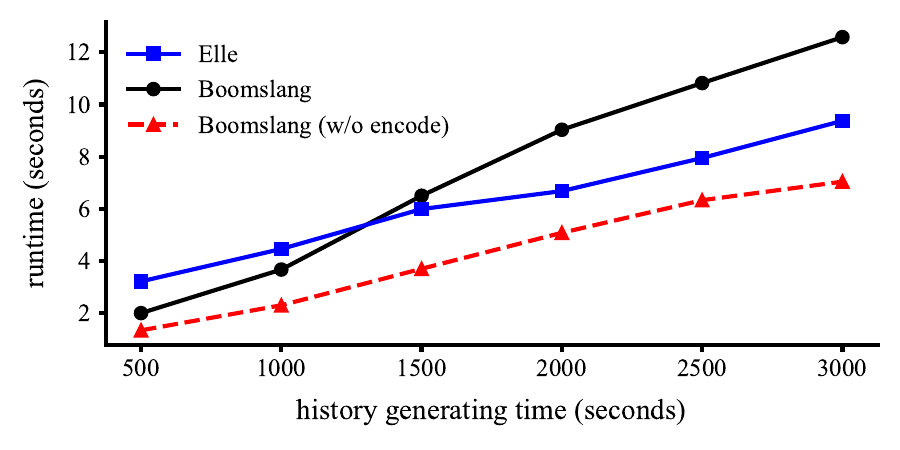}
    \caption{Checking serializability: \sys and Elle.
    }
    \label{fig:compare-with-elle}
\end{figure}
 In this experiment,
we run Jepsen's ListAppend-J benchmark and vary the history collection time,
which is approximately proportional to the number of transactions.
We compare the checking time between \sys and Elle.
We adopt 24 sessions.
Figure~\ref{fig:compare-with-elle} shows the results. 
The performance is comparable, yet \sys has a greater slope than Elle.
This is expected:
\sys uses SMT solvers and hence has to pay an extra cost for encoding.
If we exclude \sys's encoding time of calling SMT solver APIs but still keep the solving time,
\sys shows similar performance.

\begin{figure}[t]
    \vspace{-1ex}
    \centering
    \includegraphics[width=0.40\textwidth]{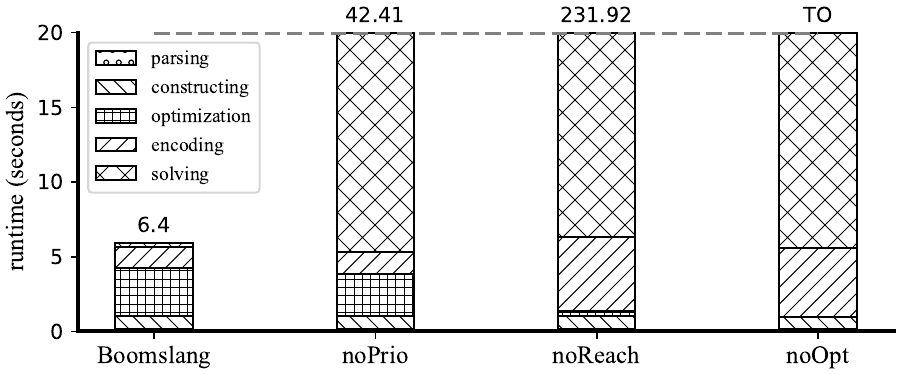}
    \caption{%
        Decomposition of \Sys runtime of BlindW-C (2K keys, 32 threads, 10K txns) under different settings.
    }
    \label{fig:breakdown}
\end{figure}
 
\headingzero{Ablation study.}
In this experiment, we conduct an ablation study
of how different phases and optimizations contribute to the total running time.
We evaluate \sys with BlindW-C benchmark on PostgreSQL under serializability.
We vary the configurations of serializability checkers
with various optimizations:
(a) \emph{\sys}: \sys with both optimizations;
(b) \emph{noPrio}: \Sys serializability checker without \optone;
(c) \emph{noReach}: \Sys serializability checker (\S\ref{ss:encoding}) without \opttwo.
(d) \emph{noOpt}: \Sys serializability checker with no optimizations.
    We measure the runtime of \sys's different phases: %
    parsing, %
    constructing, %
    optimization, %
    encoding, %
    and solving. %
    Figure~\ref{fig:breakdown} shows the results.
    \Sys times out (>600s) without optimizations.
    With \optone, \sys (the noReach bar) becomes solvable because \optone has a
    better guided search in the solving process.
    With \opttwo, \sys (the noPrio bar) is faster than the noReach bar because many \superposs have been pruned.

%
%
%
%
%
%
%
%
%
%
%
%

\begin{figure}[t]
    \centering
    \includegraphics[width=0.3\textwidth]{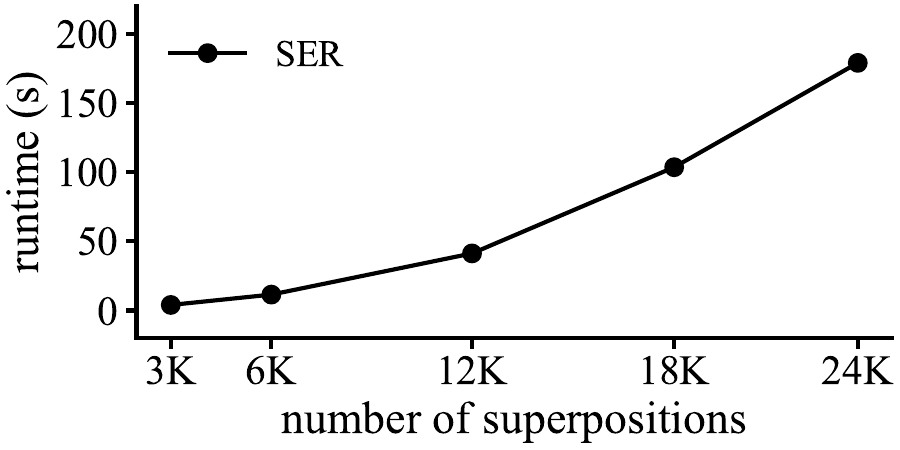}
    \caption{Checking \ser: runtime grows superlinearly with the number of \superposs.
     }
    \label{fig:perf-superpositions}
\end{figure}

%
\headingzero{Runtime as a function of \#\superposs.}
In this experiment, we study the relationship between the checking time and the number of superpositions.
Figure~\ref{fig:perf-superpositions} shows the results.
The checking runtime grows superlinearly,
which is expected as the \ser checking is an NP-Complete problem~\cite{tan20cobra}.

\headingzero{Detecting real-world violations.}
To confirm \sys's ability to catch real-world anomalies,
we experiment with 15 real-world \traces
that have known isolation level violations
and three \traces we collected from Tapir.
We download the 15 \traces from online repositories~\cite{cobralogs,plumegit,verisogit,li2025-git,jepsen-site}.
Additionally, we generate three new \traces for Tapir by running
a slightly modified BlindW-T benchmark on a single machine.
These \traces cover four isolation levels: serializability, snapshot isolation,
strict serializability, and PL-2+.
Figure~\ref{fig:real-world-violations2} shows the results.
\Sys detects all the violations within 26s.

\begin{figure}
	\scriptsize
	\begin{tabular*}{\columnwidth}{@{\extracolsep{\fill}} llllll @{}}
		\toprule
		\textbf{Iso.} &  \textbf{Anomaly} &  {\textbf{Database}}   & {\textbf{\#Txns}} &
		{\textbf{Time}}   \\
		\midrule
		\multirow{9}{*}{SER}  & G2-anomaly~\cite{yuga-g2} &  YugaByte$^\dagger$ 1.3.1.0 & 37.2K &  1.6s  \\
		& Lost updates~\cite{yuga-disw} &  YugaByte 1.1.10.0 & 2.8K & 25.55s   \\
		& G2-anomaly~\cite{cock-g2} &  CockroachDB beta.  & 446 & 0.08s   \\
        & Read uncommitted~\cite{cock-blog} &  CockroachDB 2.1 & 20$^\star$ & 0.05s   \\
		& Read skew~\cite{fauna-page} &  FaunaDB 2.5.4 & 8.2k & 14.05s   \\
		& Fractured reads~\cite{liu2025veriso} & Tapir & 547 & 0.07s \\ %
		& Fractured reads~\cite{liu2025veriso} & Tapir & 480 & 0.06s \\ %
		& G1-anomaly~\cite{adya99weak} & Tapir & 13156 & 0.68s \\ 
		& G2-anomaly~\cite{adya99weak} & Tapir & 9082 & 0.66s \\ 
		\hline
		\multirow{8}{*}{SI}  & Lost updates~\cite{mongodb-violation} &  MongoDB 4.2.6 & 23.2K & 0.28s \\
		& Aborted read~\cite{mongodb-violation} &  MongoDB 4.2.6 &  2.2K & 0.08s  \\
		& G1c-anomaly~\cite{mongodb-violation} &  MongoDB 4.2.6 &  1.1K &  0.07s \\
		& Read future writes~\cite{mongodb-violation} &  MongoDB 4.2.6 & 4.6K & 0.14s   \\
		& Read skew~\cite{tidb-read-skew} &  TiDB 2.1.7 &  9.1K & 0.2s   \\
		& Non-repeatable read~\cite{Eisenberg2000} & MariaDB$^*$ 10.4.22 & 1K & 0.16s \\ %
		& G-SI-anomaly~\cite{adya99weak} & Yugabyte 2.11.1 & 20 & 0.05s \\ %

		\hline
		SSER & Time inversion~\cite{lu23ncc} & Tapir & 68 & 0.05s \\
		\hline
		PL-2+ & Thin-air read~\cite{jeremy2005} & AntidoteDB 0.2.2 & 120 & 0.03s \\ %
		\bottomrule
	\end{tabular*}
	\caption{\sys detects real-world violations.
	``CockroachDB beta.'' represents CockroachDB-beta 20160829.\\
	$^\star$ The report only contains a small fragment of the original \trace. \\
	$^*$ This is short for MariaDB-Galera. \\
	$^\dagger$ This is short for YugaByteDB.
	}
	\label{fig:real-world-violations2}
\end{figure}

\section{Related work}
\label{s:relwork}

\headingzero{Database \checkers.}
The closest related works to \sys are
Cobra~\cite{tan20cobra},
PolySI~\cite{huang2023efficient},
Viper~\cite{zhang2023viper},
and Emme~\cite{clark21verifying}.
\sys subsumes their encodings and performs similarly or better than their
original implementations.
dbcop~\cite{biswas19complexity,dbcop}
verifies multiple isolation levels, like \sys.
The algorithm complexity is $O(n^c)$
where $n$ is the number of transactions
and $c$ is the number of sessions.
So, it does not scale well.
Elle~\cite{kingsbury20elle} is a versatile
\checker. It verifies multiple isolation levels
and has several ``modes'', some of which are not sound.
It leverages atomic operations such as ``append''
to reveal \vorder of writes.
However, Elle cannot support arbitrary user workloads.
Ouyang's checker~\cite{ouyang21verifying}
is a white-box SI checker.
Compared with these \checkers,
\sys adapts to more configurations,
supports all operations, and allows arbitrary user workloads.

%

\headingzero{Testing and bug finding in concurrent databases.}
A body of work focuses on detecting bugs under different isolation levels,
including recent systems such as TxnCheck~\cite{jiang2023detecting},
Troc~\cite{dou2023detecting}, and Leopard~\cite{li2023leopard}.
Unlike the \checkers discussed earlier, these bug-finding systems trade completeness or
soundness for performance and may instrument user workloads to expose database
states.
Similarly, another research direction evaluates the consistency levels that
users can expect from cloud storage
systems~\cite{zellag12consistent,wada11data,rahman12toward,lu15existential,kim15caelus}.
In contrast, \sys has a different focus. It does not
modify user applications,
and it provides results that are complete and sound.

\headingzero{Other consistency checkers.}
Beyond database checkers, there exist checkers for other consistency models,
such as linearizability~\cite{burckhardt2010line,lowe2017testing} and memory
models~\cite{torlak2010memsat}. At a high level, \sys addresses a similar
problem: searching for a valid sequence. However, \sys operates in a different
context---transactions---which group multiple operations under varying isolation
guarantees defined by different isolation levels. Furthermore, \sys unifies all
configurations within a single framework.

\frenchspacing

\begin{flushleft}
  \footnotesize
  \setlength{\parskip}{0pt}
  \setlength{\itemsep}{0pt}

  \renewcommand{\UrlFont}{\rmfamily}
  \bibliographystyle{abbrv}
  \bibliography{conferences-long-with-abbr,bibs}
\end{flushleft}

\appendix
\clearpage
\section{Appendix}

\subsection{Graph-based isolation level definition}

\headingzero{Serialization graph.}
Most existing \checkers~\cite{tan20cobra,huang2023efficient,zhang2023viper,clark21verifying,clark2024validating,kingsbury20elle}
build upon the serialization graph related theories.
In a serialization graph, each node represents a
committed transaction, while edges denote dependencies between different transactions. 
Conflict dependencies are the most common type of dependencies; they represent the database's scheduling of two transactions accessing the same key.
Based on operation types (i.e., read or write), there are three conflict dependencies: 
\label{ss:dependency}
\begin{myitemize2}
    \item \emph{\wrrel}: a transaction $T_j$ reads the value written by transaction $T_i$ ($T_i \wrarrowx T_j$).
    \item \emph{\wwrel}: $T_i$ writes a key, and $T_j$ overwrites it ($T_i \wwarrowx T_j$).
    \item \emph{\rwrel}: $T_i$ reads a value that is later overwritten by $T_j$ ($T_i \rwarrowx T_j$).
\end{myitemize2}

\heading{Predicate dependency.}
Beyond basic single-key reads and writes, we can have finer-grained conflict dependencies for predicate-based queries---such as range queries
and \iters---conflict with writes and deletions when their updated key falls within the
queried range.
For example, a deletion \ttt{del(y)} in $T_i$ deletes the key \ttt{y} $\in$ \ttt{[x,z)}
(assuming an alphabetical order on keys),
which conflicts with a range query \ttt{scan(x,z)} in $T_j$.
For transactional key-value stores,
there are two types of predicate dependencies:
\begin{myitemize2}
    \item \emph{\pwrrel}: $T_j$ contains a query with predicates and the query reads the
        key-value pair written by transaction $T_i$ ($T_i \pwrarrowx T_j$).
    \item \emph{\prwrel}: $T_j$ performs a deletion (or write) which deletes (or inserts)
        a key-value pair observed (or unobserved)
        by a predicate-based query in $T_i$ ($T_i \prwarrowx T_j$).
\end{myitemize2}

\noindent
In addition to conflicting dependencies, %
other types of dependencies also exist.
For example, \emph{start-dependency} represents that transaction
$T_j$ begins after $T_i$ commits ($T_i \to T_j$). 

An isolation level can be defined by rules on the serialization graph, namely Adya's framework~\cite{adya99weak,adya00generalized}.
As an example, \Ser~\cite{bernstein79formal,papadimitriou79serializability,weikum01transactional}, the gold-standard isolation level,
requires that concurrent transactions execute as if they were run
sequentially on a single-copy database.
A trace is serializable if and only if its serialization graph is acyclic with conflict dependencies.
As another example, for snapshot isolation, a serialization graph must not contain any cycle with exactly one
anti-dependency edge~\cite[G-SIb]{adya99weak}.

Because these graph-based definitions are straightforward formalism of
isolation levels, an intuitive checking approach is to construct the
serialization graph with the trace and look for violations to check if the
promised isolation is met.
Note that this requires obtaining all dependency information to construct the edges,
which envelope tracing does not provide.

\subsection{Abstract Semantic Graph}

Below, we elaborate on each of the three components of ASG.
First, the known graph captures the dependencies identifiable from the trace.
The structure of this graph depends on the target isolation level. For example,
when checking serializability, nodes represent committed transactions and edges
represent all five types of dependencies (\S\ref{ss:dependency}).
This yields a serialization graph, which may be incomplete due to unknown information in the trace.
In contrast, for snapshot isolation, the graph follows the definition of
BC-graphs~\cite{zhang2023viper}, where nodes correspond to transaction begin or commit events.
When checking read committed, only three types of dependencies are included;
\rwrels and \prwrels are excluded~\cite{adya99weak,adya00generalized}.

Second, the \writeinfo captures updates in the trace:%
\[
Map\{ \t{Key} \to \{\varwset: Set\langle id,val\rangle , \varvo: Set\langle id, id \rangle\}  \}.
\]%
This includes $\varwset$, the set of all writes and deletes to a key with their transaction id and value,
and, if available, $\varvo$---version order, a partial order of writes updating the same key (in Adya's formalism, a version order should be a total order; we allow this total order to be partially known).

Finally, the \readinfo summarizes all retrieval operations using a map:
\vspace{-.5ex}
\[
Map\{(\t{Key},\t{Val}) \to Set\langle [r\_id, \varpotentialw, \t{type}, \t{verSet}] \rangle \}.
\]
\vspace{-.5ex}
Each retrieved result, indexed by keys and values, maps to one or more
retrieval operations along with their details:
$r\_id$ identifies the transaction performing the operation;
$\varpotentialw$ lists all candidate writes this operation may read from;
$\t{type}$ indicates the operation type (i.e., read, range query, or iterator);
and, if available, $\t{verSet}$ provides the version set~\cite{adya99weak},
auxiliary information for range queries and iterators.

\heading{Constructing ASG.}
\Sys begins by scanning the trace to extract all committed transactions,
creating the known graph based on the targeted isolation level.
Next, it identifies all writes and deletes,
storing their information in \writeinfo.
Finally, \sys analyzes the retrieval operations to construct \readinfo.
If a retrieved value does not correspond to any written value,
\sys detects a data integrity violation and rejects the trace.
\Sys also automatically infers dependencies when possible
and incorporates user hints (described in \S\ref{ss:usermodules})
to generate dependencies.

\label{ss:handlingpredicate}
\sys handles predicate-based queries (i.e., range queries and \iter)
differently depending on the assumptions and setup.
If version sets~\cite{adya99weak} are available (e.g., in a white-box setup), \sys uses
the version sets to construct dependencies for range queries and iterators.
When version sets are absent---such as in a black-box setup---but \tombstones (i.e., logically deleted items that remain visible to queries)
are assumed, all keys in the queried range are returned, including both existing and deleted ones.
For existing keys, \sys analyzes the returned values to establish \wrrels;
for deleted keys, their \tombstones are returned, revealing the transactions responsible for the deletions.
This allows \sys to establish \wrrels to the deleting transactions.
If \tombstones are not used, the trace contains \unknowns,
requiring \superpos to handle such cases, which is described
in Section~\ref{ss:predicate}.

\heading{Other dependencies.}
\Sys also supports dependencies beyond those defined by isolation levels (\S\ref{ss:dependency}),
such as session order and real-time order.
Session order requires that transactions issued within each session
(e.g., by each client) must respect their issuing order.
Users can implement a \sys module to add session-order edges to the ASG.
Similarly, real-time order can be supported by introducing real-time edges between transactions:
if one transaction commits earlier (based on wall-clock time) than
another begins, the former must precede the latter.
More generally, any ``happened-before'' requirements---like monotonic reads/writes,
bounded staleness, and consistent prefix---can be represented using similar
approaches in \sys.

\subsection{\sys's IR}

\begin{figure}
\lstdefinelanguage{PythonHighlighted}{
  keywords={def, return, for, in, if, else, global},
  basicstyle=\ttfamily\scriptsize,
  keywordstyle=\color{blue}\bfseries,
  morekeywords=[2]{ASG, genIR, genReadDep, genSessionOrder, genSessionDep},
  keywordstyle=[2]{\color{purple}\bfseries},
  sensitive=true,
  comment=[l]{\#},
  commentstyle=\color{green!60!black},
  stringstyle=\color{red},
  morestring=[b]',
  morestring=[b]"
}
\begin{lstlisting}[language=PythonHighlighted,
                   numbers=left,
                   frame=lines,
                   framesep=3pt,
                   %
                   ]
IR ::= {
    graph           : Graph,
    expressions     : Set<LogicExpr>,
    superpositions  : Set<Superposition>
}

Graph ::= {
    vertices : Set<Node>,
    edges    : Set<Edge>,
    edgeVar  : Map<Edge, BoolVar>
}

Node ::= TransactionID | OperationID

Edge ::= (src : Node, dst : Node, type : EdgeType)

EdgeType ::= RW | WW | WR | Session | ...

LogicExpr ::= BoolConst
            | BoolVar
            | LogicExpr AND LogicExpr
            | LogicExpr OR  LogicExpr
            | NOT LogicExpr
            | ...

Superposition ::= Set<LogicExpr>
\end{lstlisting}
\caption{\sys's IR.}
\label{fig:algo:irdef}
\end{figure}
 
Figure~\ref{fig:algo:irdef} defines \sys's IR.

\subsection{Building \superposs}

\heading{Common \superposs and logical expressions.}
\sys distinguishes itself from prior checkers through
its generalizability, enabled by its support for the low-level IR.
In the following, we introduce two common types of \superposs and three
classes of \irimplies,
some of which capture previously uncheckable conditions.

The first \superpos handles black-box traces when \vorder is unknown.
The construction reads as:
for each pair of conflicting transactions $T_i$ and $T_j$ that updates the same key $x$,
they form a \superpos of $\langle T_i \wwarrowx T_j, \ T_j \wwarrowx T_i \rangle$.
In addition, all \superposs collectively should
provide a total order of writes to the same key.

Second, \sys supports \tval,
which can happen when
there are duplicated writes that store the same value for key $x$.
Then, reading such a value has multiple possibilities.
The read transaction, $T_r$, may read from any one of the write transactions,
from $T_{w1}$ to $T_{wn}$.
For such \unknowns,
a \superpos is constructed as:
\[
    \langle T_{w1} \wrarrowx T_r,\ \dots,\ T_{wn}
    \wrarrowx T_r \rangle.
\]

\label{ss:irimplies}
The first type of \irimplies captures another type of unknown scenario:
a set of unknown \wrrels from potential write transactions
$T_{w{1}}, \cdots, T_{w{n}}$ to read transaction $T_r$: $T_{w{1}} \wrarrow{x} T_r,\ \dots,\ T_{w{n}} \wrarrow{x} T_r$ and a set of unknown \wwrels
$\{T_{wj} \overset{\text{?}}{\leftrightarrow} T_{p_1}, 
T_{wj} \overset{\text{?}}{\leftrightarrow} T_{p_2},
\dots\}$
where $1 \le j \le n$, and $T_{p_1}, T_{p_2},\ \dots$ have conflicting writes to $x$.
Then, the \rwrel is unknown.
The corresponding \irimply is defined as:
\[
    (T_{wj} \wwarrow{x} T_{p_l} \land 
    T_{wj} \wrarrow{x} T_r)
    \Rightarrow T_r \rwarrow{x} T_{p_l}
\]

The second type of \irimplies is a simplified case if a \wrrel $T_{wj} \wrarrow{x} T_r$ is known.
The corresponding \irimplies can be expressed as:
\[
    T_{wj} \wwarrow{x} T_{p_l} \Rightarrow T_r \rwarrow{x} T_{p_l}
\]

Similarly, another simplified case occurs if $T_{wj} \wwarrow{x} T_{p_l}$ is known.
The corresponding \irimplies is:
\[
    T_{wj} \wrarrow{x} T_{r} \Rightarrow T_r \rwarrow{x} T_{p_l}
\]

\label{ss:predicate}
\heading{\CF{\superpos} for predicate-based queries.}
As mentioned earlier (\S\ref{ss:handlingpredicate}),
black-box traces without \tombstones contain \unknowns for predicate-based queries,
such as range queries and iterators.
To tackle such \unknowns,
\sys does the following:
for each key $x$ that satisfies the range and appears in the \trace but is not
returned by the range query or the \iter (in $T_i$),
if the initial state does not contain $x$,
then \sys creates a \superpos:
\[
    \langle T_{init} \pwrarrowx T_i,\ T_{del1} \pwrarrowx T_i,\ \cdots,\ T_{deln}
    \pwrarrowx T_i \rangle,
\]
where $T_{init}$ is the abstract initial transaction,
and $T_{del1}$ to $T_{deln}$ are all transactions that delete key $x$. Otherwise, \sys creates a \superpos:
\[
    \langle T_{del1} \pwrarrowx T_i,\ \cdots,\ T_{deln}
    \pwrarrowx T_i \rangle,
\]
This \superpos also applies to read operations that return a \mnull value, as
it remains
ambiguous whether the key has been deleted---and if so, by which
transaction---or if it has not yet been inserted.

\subsection{Experiment configuration details}

\heading{Setup.}
All experiments are run on a 12-core
AMD Ryzen 5900 machine,
with 64GB of memory and Ubuntu 20.04.
But we configured the checker to use at most 40GB.

\heading{Configurations.}
We run RUBiS~\cite{rubis}, Twitter~\cite{twitter}, TPC-C~\cite{tpcc},
BlindW-C~\cite{cobra_verifier,cobra_bench},
on PostgreSQL 15.2~\cite{postgresql_website}, YugaByteDB 2.25.1~\cite{yuga_git},
CockroachDB 25.1.4~\cite{cockroach_paper};
We run RandomBench, and RandomRMWRange on TiKV 7.5.6~\cite{tikv}.
We run RandomRange on TiDB 8.5.4.
We additionally run the \blindwt~\cite{zhang2015} and Retwis~\cite{zhang2015} on Tapir.
We run 24 concurrent clients by default.

\heading{Details of our benchmarks.}

\begin{myitemize2}
    \item \emph{RandomBench}:
    this microbenchmark includes all the operations that we support, 
    including Get, Put, range queries, true deletions and iterators.
    No prior black-box \checkers support this benchmark.
    Written values are non-unique,
    and each transaction has 8 operations:
    25\% reads, 25\% writes, 5\% deletions, 10\% range queries, 10\% \iters and 25\% read-modify-writes.
    Range queries retrieve all key-value pairs within
    a specific key range specified by [$k_1, k_2$),
    where $k_1$ is generated randomly in [$1, NUM\_KEYS$],
    and $k_2$ is set to be $k_1 + RandomInt(1, 100)$.
    Each \iter queries top-k items
    within a specific key range specified by [$k_1, k_2$),
    where $k_1, k_2$ are generated the same way as RandomRange
    and the k is randomly picked from $[1, 10]$
    and unknown to the \checker.

    \item \emph{RandomRange}:
    this microbenchmark evaluates range queries.
    No prior black-box \checkers support this benchmark.
    Written values are non-unique,
    and each transaction has 8 operations:
    25\% reads, 25\% writes, 5\% deletions, 20\% range queries and 25\% read-modify-writes.

    \item \emph{RandomRMWRange}: this benchmark simulates Emme's experiments~\cite{clark2024validating}. %
    Clients issue transactions with 8 operations, each being a read, a write or a range query.
    Any write is preceded by a read of the same object, forming a read-modify-write. %
    Any range query is preceded by a global range query to recover the version set.
    The probabilities of read-modify-writes and range queries are 90\% and 10\% respectively.
    Reads and writes pick a key randomly from 10K predefined keys,
    while range queries select key pairs in the same way as RandomRange and RandomBench.
    The written values in RandomRMWRange are unique to infer the version order.

\end{myitemize2}

We also implemented our own 
BlindW
(distinguished from Cobra's BlindW-C and Tapir's BlindW-E) benchmark that is
the same as RandomBench except the ratios of reads and writes are set to 50\%
each. 

\subsection{More experimental results}
In this section, we study the scalability of \sys in terms of both checking time and memory cost.

\headingzero{Applications.}
We experiment with the existing 
BlindW-C~\cite{cobra_verifier,cobra_bench} from Cobra~\cite{tan20cobra}.

\headingzero{\sys's scalability}
In this experiment, we study how
performance varies across different isolation levels,
and how \sys scales as the problem size grows.
We use BlindW-C because scalability testing with basic reads and writes
is a widely adopted method in this field.

For the setup,
we execute BlindW-C on PostgreSQL with varying numbers of transactions and 10K keys under serializability.
As serializability is the strongest isolation level, the collected \traces
also satisfy snapshot isolation and read committed.
We then use \sys to verify serializability, snapshot isolation and read committed
for the collected \traces.

\begin{figure}[t]
    \vspace{-1ex}
    \centering
    \includegraphics[width=0.35\textwidth]{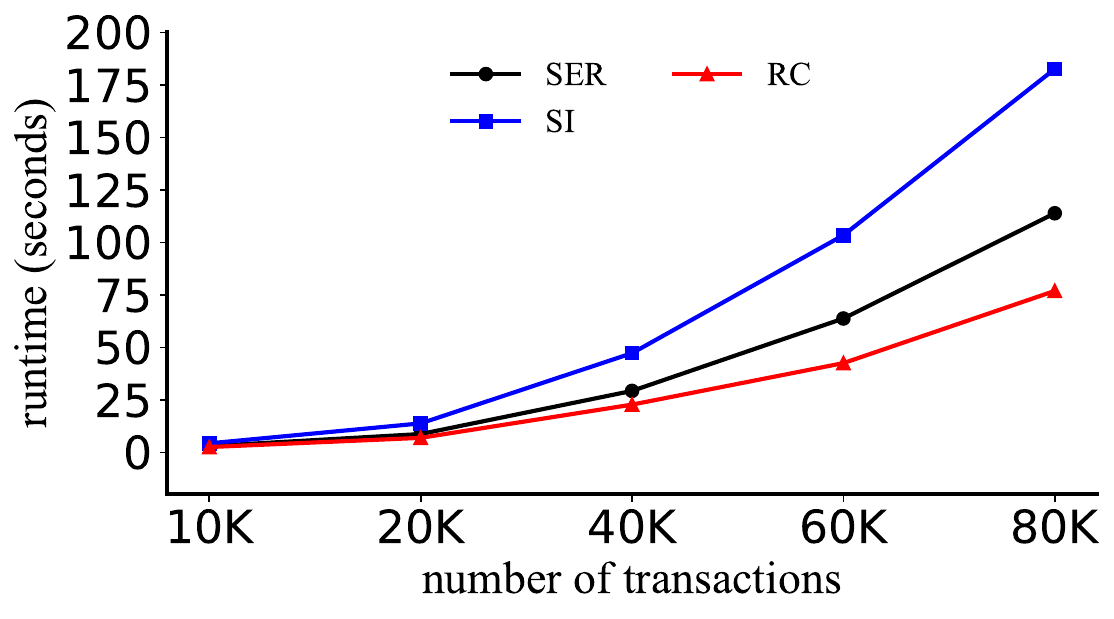}
    \caption{\sys's scalability on checking serializability (SER), snapshot isolation (SI), and read committed (RC).
     }
    \label{fig:scalability}
    \vspace{-2ex}
\end{figure}

Figure~\ref{fig:scalability} shows the runtime results.
\Sys can handle up to 80K transactions,
after which we stop due to memory limitations.
Among the isolation levels,
read committed is the fastest to verify because it only considers \wwrels and \wrrels, excluding \rwrels.
Snapshot isolation and serializability require consideration of all types of dependencies.
Snapshot isolation is the most time-consuming because its BC-polygraphs contain twice as many nodes and
more intra-transaction edges compared to serializability.

\begin{figure}[t]
    \vspace{-1ex}
    \centering
    \includegraphics[width=0.35\textwidth]{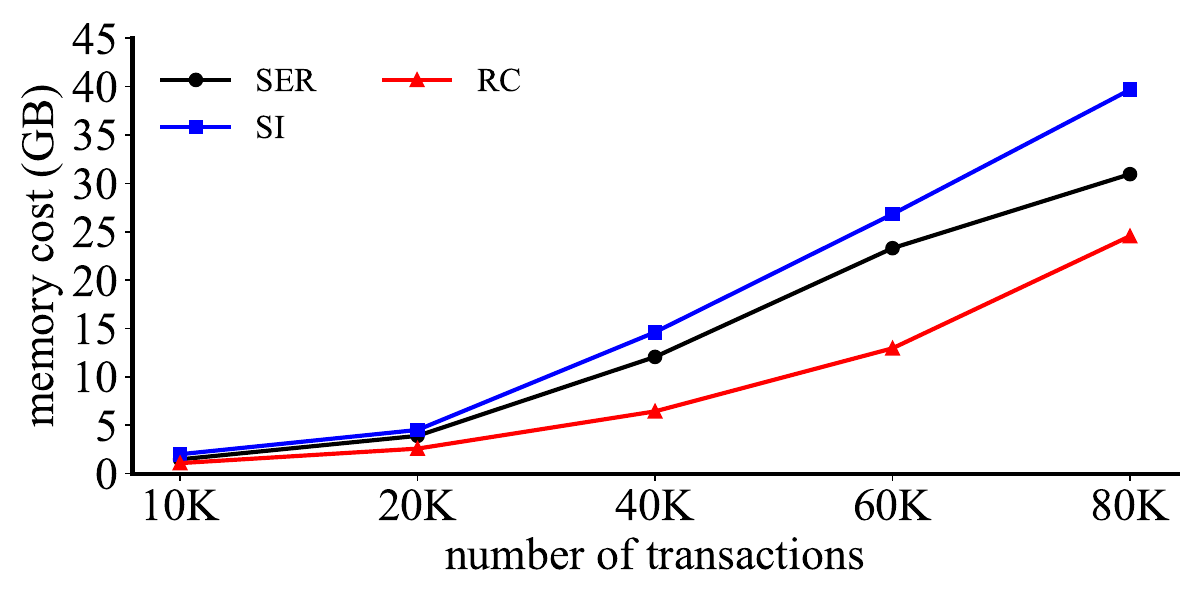}
    \caption{\sys's scalability on checking serializability (SER), snapshot isolation (SI), and read committed (RC).
     }
    \label{fig:memory-scalability}
    \vspace{-2ex}
\end{figure}
 Figure~\ref{fig:memory-scalability} shows the memory cost (max resident set memory size) as the number of transactions increases.
The memory cost grows superlinearly.
Snapshot isolation uses double graph nodes as serializability, has extra intra-transaction edges than serializability, and contains the same set of superpositions as serializability, so it consumes the most memory.
The memory cost achieves around 40GB when checking SI for the 80K trace.
Checking serializability involves all the types of dependencies while checking read committed only contains non-anti dependencies. They share the same set of graph nodes. So checking read committed costs less memory than checking serializability.

\begin{figure}[t]
    \centering
    \includegraphics[width=0.45\textwidth]{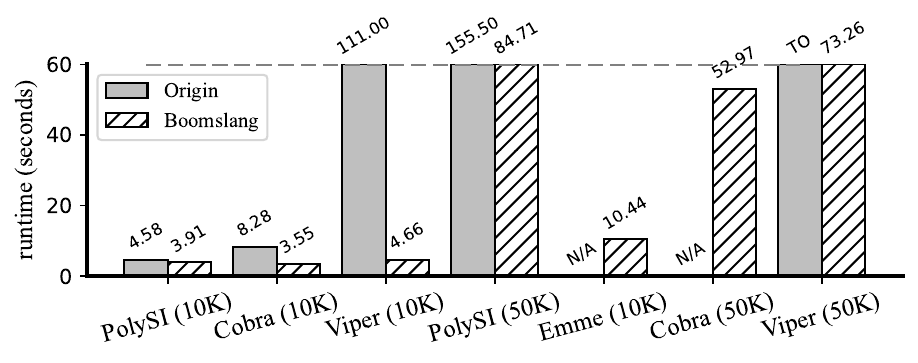}
    \caption{Comparison of encoding implementations.
    ``Origin'' refers to the original implementation,
    while ``Boomslang'' represents \sys's implementation of the same encoding.
    The x-axis shows the encoding and the \trace size;
    for example, ``Cobra (10K)'' indicates Cobra's encoding for checking serializability
    on 10K transactions.
    All \traces are generated from BlindW, except Emme's, which is generated using RandomRMWRange.
    }
    \label{fig:encodings}
    \vspace{-2ex}
\end{figure}
 \headingzero{Same encoding, different implementations.}
We evaluate the same encodings implemented in both \sys and their original systems.
The results show that \sys's generalizability incurs modest performance overhead.
Figure~\ref{fig:encodings} shows the results for BlindW-C 10K and 50K transactions.
\sys runs faster than all of its counterparts
because \sys has additional shared optimizations, such as \optone;
and \sys uses Java, whereas Viper uses Python.

\end{document}